\documentclass[onecolumn]{aa} 
\usepackage{graphicx}
\usepackage{txfonts}
\usepackage{color}

\usepackage{natbib}
\bibpunct{(}{)}{;}{a}{}{,} 
\usepackage{graphicx}
\usepackage{lscape}
\usepackage{rotating}
\usepackage{comment}

\def\simle{\lower 2pt \hbox {$\buildrel < \over {\scriptstyle \sim }$}}
\def\simge{\lower 2pt \hbox {$\buildrel > \over {\scriptstyle \sim }$}}

\begin{document}

\title{Predicted power in ultra high energy cosmic rays from active galaxies}

\author{Lauren\c{t}iu I. Caramete
	\inst{1,2,}\thanks{Member of the International Max Planck Research School
(IMPRS) for Astronomy and Astrophysics at the Universities
of Bonn and Cologne}
		 \and Oana Ta\c{s}c\u{a}u\inst{3}
		 \and Peter L. Biermann\inst{1,} \inst{4,} \inst{5,} \inst{6,} \inst{7}
		 \and Todor Stanev\inst{8}
		}
\institute{Max Planck Institute for Radio Astronomy, Auf dem H\"ugel 69, 53121 Bonn, Germany
	\and
Institute for Space Sciences, P.O.Box MG-23, Ro 077125 Bucharest-Magurele, Romania
\and
Department of Physics, University of Wuppertal, Gauss-Str. 20, D-42119 Wuppertal, Germany
\and
Karlsruhe Institute of Technology, P.O. Box 3640, 76021 Karlsruhe, Germany	
\and
Department of Physics and Astronomy, University of Alabama, Tuscaloosa, Alabama 35487, USA
\and
Department of Physics, University of Alabama at Huntsville, AL 35899, USA
\and
Department of Physics and Astronomy, University of Bonn, Endenicher Allee 11-13, Bonn, Germany
\and
Bartol Research Institute, University of Delaware, DE 19716, USA
}

%
%
%
%
%
\abstract
 {As more and more data are collected by cosmic ray experiments such as the
 Pierre Auger Observatory and Telescope Array (TA), the search for the sources of the Ultra High
 Energy Cosmic Rays (UHECR) continues. Already we have some hints about the sources or type of sources involved and more work is required to confirm any of this.}
 {We intend to predict the UHECR fluxes and the maximal energies of particles
 from two complete samples of nearby active galaxies, selected at radio
 and far-infrared frequencies. Also, we investigate the magnetic scattering of
 the UHECR path in the intervening cosmic space.}
 {We propose here a new method of searching for the sources of the UHECR
 in three steps, first we model the activity of the type of sources and
 get the flux of UHECR and a maximal energy for particle acceleration,
 then we model the interaction and angle deflection in the intergalactic
 space and finally we simulate the distribution of the cosmic rays events
 that can be statistically compared with future data of the cosmic rays
 observatories.}
 {We analyzed two classes of sources, gamma ray bursts (GRBs) and Radio Galaxies (RGs). Ordering by the UHECR flux, few RGs are viable candidates, as for GRB many sources are viable candidates, requiring less scattering of the particles along their path to Earth to interpret the presently observed sky distribution. Most of the flux from RGs comes from the Southern sky, and most of the flux of particles from GRB comes from the North, although the differences are so small as to require large statistics to confirm this. The intergalactic and
 Galactic magnetic fields may help to distinguish the two extreme cases, also pure protons from heavy nuclei at the same energy.
 The simplest model may be that Cen A produces nearly all of the observed UHECR, if they are mostly heavy nuclei, and M87 if they are mostly light nuclei. As a consequence flat spectrum radio sources such as 3C279 should confirm the production of UHECR in this class of sources through energetic neutrinos.}
{}
 \authorrunning{L.I. Caramete et al.}
 \titlerunning{Predicted power in ultra high energy cosmic rays
 from active galaxies}
   \maketitle

\section{Introduction}

 The question of where the ultra high energy particles originate,
 is still unresolved. \citet{1963SvA.....7..357G} predicted that
 the radio galaxies Cen A (= NGC5128), Vir A (= NGC4486 = M87), and
 For A (= NGC1316) should be good candidates to provide most of the
 extragalactic cosmic rays. The case for RGs was made
 repeatedly since then, in many papers, e.g.
 \citet{1987ApJ...322..643B,1993cac..book...12B,2002PhRvD..65b3004I}.
 The topic was treated in depth in the book by \citet{2010hecr.book.....S},
 and review articles by
 \citet{2011arXiv1101.4256K,2010arXiv1011.1872S,2009APh....31..138B}.
 The new Pierre Auger Observatory data,\
 \citet{2007Sci...318..938T,2008APh....29..188P,2010APh....34..314T},
 suggest that the directions of the incoming events are not isotropic
 but cluster like the distribution of matter in the cosmos around
 different regions of the observed sky. It will be interesting to check 
this with present and future data and also with observations taken from the Northern hemisphere like,
 HiRes observations \citep{2008APh....30..175T,2010ApJ...713L..64A}.
 Whether these events are due to gamma ray bursts in starburst galaxies,
 intergalactic shock waves, active galactic nuclei or some other
 phenomenon within galaxies, we do not know yet, the present data
 are too scarce to select the true sources.

 As there are several hypotheses in the literature, it becomes important
 to test different ideas using complete samples of classes of candidate
 sources.  Such samples need to be flux density limited, and redshift
 limited, in order to relate them to studies made in the last few decades.

 Here we propose to test two hypotheses, long under discussion,
 a) that Gamma Ray Bursts are the dominant sources, and
 b) that Radio galaxies are the dominant sources, in both cases
 using complete samples. In addition, RGs pointing their
 relativistic jet towards Earth are observed as flat spectrum radio
 sources, and so provide a test through neutrino observations.

Observationally, complete samples can be constructed with a flux density limit and with a redshift limit, and with information, whether the sample is isotropically distribution in their symmetry angles like the jet direction of the central super-massive black hole, or their overall spin-axis of the host galaxy. This redshift limitation throws up a very serious problem: As we have to expect that the ultra high energy cosmic ray flux is dominated by a huge number of very distant sources, for which scant redshift information is available, how can such a limited sample be representative? The answer can be obtained in three steps: First, within the first complete sample, with a redshift limit, we can ask, whether the fainter sources contribute more than the first one or two sources. If the fainter sources contribute a negligible flux, then we proceed to ask the question again in a statistical sense, using a radio galaxy luminosity function without a redshift limit, \citet{2001MNRAS.322..536W}, and again ask, whether the fainter sources contribute significantly. Third, we can propose a test using secondaries, that do not interact with either photon fields or magnetic fields, flat spectrum radio sources, which are normal usually low power radio galaxies, that have their relativistic jet pointed at us, and so are visible to sub-TeV gamma rays to a substantial redshift, and to TeV neutrinos depending on flux to any redshift. The method then is to calculate a maximal flux at Earth, so that any possible interaction with either photon fields or magnetic fields would lower the flux detectable at Earth; if already the maximal flux shows a clear result, than a fortiori any interaction would lower the flux detectable at earth and so would strengthen our result. It turns out that we obtain a consistent answer.

When we consider these three complete samples, we emphasize that for charged particles we require isotropically oriented samples, and for neutral particles we require sources strictly oriented towards Earth.  It has been shown repeatedly that steep spectrum radio galaxies are isotropically oriented to a very good approximation, and flat spectrum radio sources are rather strictly oriented towards Earth (e.g. \citet{1984AJ.....89..323G}).  Similarly, for far-infrared selection a complete sample is very nearly isotropically oriented (e.g. \citet{1996ARA&A..34..749S}) and so fulfills the same condition for GRBs.

\section{Intergalactic scattering model}

 Basic questions of the effect of intergalactic and galactic magnetic
 fields on the propagation of ultra high energy charged particles
 are whether a) is there (almost) no effect, b) a general systematic shift in the angle of arrival directions of incoming events or c) is there a general scattering
 \citep{2008ApJ...682...29D}.  To have no effect is less plausible
 as we do have magnetic fields in the galaxies and in the intergalactic
 space that can be quite large in scale. Any systematic shift is not
 clearly apparent at the present time with the sparse data, while a
 general scattering seems required.  There may be a general systemic
 shift, which would provide a location-dependent change of direction
 for the least scattered events from any source.  For lack of strong
 evidence we ignore such a plausible shift for the moment, although
 a systematic shift in events
 has been speculated to be a contributing reason, as possibly due to
 the Galactic magnetic wind with a symmetry with respect to the
 galactic structure
 \citep{
2010ApJ...720L.155G,2008ApJ...674..258E}.

 Regarding previos work, sky distributions were analyzed in many papers, e.g.
 \citet{2010ApJ...710.1422R,2008arXiv0805.1746S,2002ApJ...572..185A,1999astro.ph.11123A,1995PhRvL..75.3056S}, the UHECR propagation was studied and
 published also in many papers
 \citet{2010arXiv1011.1872S,2008ApJ...682...29D,2006AN....327..575D,2005PhRvD..72d3009A,2003PhRvD..68j3004S,2002PhRvD..66j3003B,2000PhRvD..62i3005S,1997ApJ...479..290S,1993A&A...273..377R,1993A&A...272..161R}. These studies are based
 on observations and simulations of the intergalactic magnetic fields
  \citep{2008Sci...320..909R,2005JCAP...01..009D,1998A&A...335...19R}.
 Also, in 1995 observations suggested a correlation with the super-galactic
 plane (\citet{1995PhRvL..75.3056S}), which contains the radio galaxies
 listed by \citep{1963SvA.....7..357G}, as other correlations were
 discussed in \citep{1996cyga.book..139B}.

 Therefore a scattering model is suggested and this will spreads arriving events
 almost evenly, the alternative to this is to have many sources, like
 hundreds of sources but no such model is currently fully convincing.
 We will discuss this issue and the limits of this statement in the following.

 A simple single scattering plasma physics approximation suggests a model
 in scattering angles of $\theta^{-2}$, per solid angle, which
 spreads events evenly into logarithmic rings $\Delta \theta / \theta \,
 = \, const.$ \citep{2011Curutiu}.  The detailed magneto-hydro-dynamic
 (MHD) simulations of \citet{2008ApJ...682...29D} support a description
 with a simple power-law for scattering angles at high energy, while at
 low energy the spreading of scattering angles is smoother and broader,
 corresponding to multiple scattering.  For simplicity we use here fits
 to the scattering laws;
 the core is to reflect what happens in the galactic disk,
 \citet{1985A&A...153...17B,1997ApJ...491..165P,1997ApJ...479..290S},
 while the general scattering distribution may reflect either scattering
 in the cosmological magnetic fields as in \citet{2008ApJ...682...29D},
 or scattering in a galactic magnetic wind halo
 \citep{1958ApJ...128..664P,1980ApJ...242...74S,1991A&A...245...79B,1993A&A...269...54B,1996A&A...311..113Z,1992ApJ...401..137P,1999astro.ph.11123A,2004ApJ...605L..33H,2005A&A...442L..49R,2006A&A...447..465C,2008Natur.452..826B,2008RPPh...71d6901K,2008ApJ...674..258E}.
 The main difference in these two sites of scattering is that we obtain
 appreciable delay times, changing the spectrum, only for scattering by
 cosmological magnetic fields \citep{2003PhRvD..68j3004S,2008ApJ...682...29D}.
 We neglect here the possibility that the source itself might be extended,
 as Cen A, with a size of 10 degree in radio data limited in sensitivity
 \citep{1993A&A...269...29J}.  We also do not take into account the effect
 of the local shear flow, the dragging of magnetic fields lines along the
 cosmological filament around Cen A with the shear flow expected to be
 parallel to the outer shape of the radio source \citep{1997ApJ...480..481K,1998A&A...335...19R,1997ApJ...477..560E,1999ApJ...511...56K,2001ApJ...560L.115G,2008Sci...320..909R}.
 This shear flow can  scatter particles, making the sites of origin
 appear correlated with the large scale filament.

 As for the propagation we realize from
 \citet{2008Sci...320..909R,2008ApJ...682...29D} that the magnetic fields
 in intergalactic space are large enough to effect large angle scattering
 even for protons. However, we have to remember one caveat here:
 In these MHD simulations, all the energy that goes into the intergalactic
 magnetic fields is derived from a non-linear structure formation driven
 by turbulent dynamo mechanism. There is no additional energy input from
 any AGN activity such as radio galaxies
 \citep{2001A&A...366...26E,2001ApJ...560L.115G,2001ApJ...560..178K}.
 Adding more energy would shift these simulations away from consistency
 given by the available observation data. Yet, when
 \citet{2001ApJ...560..178K,2001ApJ...560L.115G} use radio galaxy data,
 they obtain seemingly reasonable levels of energy input from AGN.
 How can one reconcile this discrepancy? Kronberg et al. use the
 P dV -work to estimate the energy input: P dV -work produces structure,
 and does not necessarily significantly enhance the overall magnetic
 field energy. So we arrive at the speculative concept, that the substructure
 of the IGM magnetic fields may be due to the work done by AGN, but
 that the overall energy content in magnetic fields comes mostly from
 the turbulent dynamo which derives its energy from structure formation.

 It follows, as shown in \citet{2009NuPhS.190...61B} that the scattering
 of charged particles might be reduced by this extreme substructure:
 at a given total energy content, a sheet of enhanced magnetic field
 scatters less, since the energy density scales with $B^{2}$, while
 the scattering scales with just $B$.


 In \citet{2009NuPhS.190...61B} we only did a 1D exercise to
 demonstrate this effect. We will do here a full 3D exploration
 of the effect. In a first step we assume that we have a homogeneous
 magnetic field, into which a radio galaxy blows a bubble of radius
 R, with the majority of the magnetic field energy squashed into a
 thin shell.  In a second step we will work out the effect on scattering
 by comparing the reduced length scale with the reduced Larmor radius
 of the motion of the particle.  Using a polar coordinate system with
 the symmetry axis parallel to the magnetic field we obtain for the
 magnetic field component parallel to the shock surface

\begin{equation}
B_{\parallel} \; = \; B_0 \, \sin \theta {\;} {\rm and} {\;} B_{\perp} \; = \; B_0 \, \cos \theta ,
\end{equation}

 where $\theta$ is the angle between the symmetry axis and the direction
 of the surface element. Then, assuming  that the magnetic field is
 increased by some factor $x$ (which would be $4$ under normal
 circumstances, like for a strong shock in a gas with adiabatic
 gas constant $5/3$), we obtain for the magnetic field energy density
 of this component an average of

\begin{equation}
\frac{5 x^2 B_0^{2}}{48 \pi} {\;} {\rm and} {\;} \frac{ B_0^{2}}{24 \pi} .
\end{equation}

 For the two components the sum is $({(2 + 5 x^{2}) B_0^{2}})/({48 \pi} )$.
 This is the average over a shell of thickness $R/x$, with a volume of
 approximately $ 4 \pi R^{3}/x$.  We will compare this with the total
 unperturbed energy of volume $4 \pi R^{3}/3 \cdot B_0^{2}/(8 \pi)$.
 If we assume the basic magnetic field to be the same we get an enhancement
 of the magnetic field energy density by a factor of $(2 + 5 x^{2})/(2 x)$,
 or for an example of $x=4$ a factor of $82/8 \simeq 10$. For larger
 values of $x$, as could occur in an isothermal shock, the limiting value
 would be $5 x/2$.  On average, the magnetic field is about 3 times
 stronger because of the squashing, in the simple case of $x = 4$, or,
 in the more general case $\simeq 1.6 \sqrt{ x}$.  Since in the limit
 of scattering by an angle $<< \pi$ the effective scattering angle runs
 with the relevant length scale divided by the Larmor radius this
 squashing runs as $\simeq (1.6 \sqrt{x} )/x = 1.6/\sqrt{x} < 1$.
 This will reduce the scattering, and can do it by large amounts if
 $x$ is large.  Furthermore, if we argue that the total magnetic
 field energy content is kept equal in the formation of substructure,
 then the effective scattering would be reduced even more.

 In conclusion, the substructure can reduce the scattering, in the
 limit of large enhancements of the magnetic field contrast.

For our simulations we will use the results of large scale simulations of propagation of charged particles in the intergalactic magnetic field.
We will use these scattering laws with reference for distance, since on a path form the source to us, most of the scattering will happen either close to the source or close to us, since both are part of the large scale structure; scattering close to the source, however, will not change the direction to us very much, where scattering close to us in space will change the angle of the direction of an event very much. It follows that the most plausible scattering law is one that is independent of distance for any distance smaller than the fundamental length scale of large scale structure, the length scale where distributions become homogeneous on average. We know that this scale is larger than 100 Mpc (e.g. \citet{2010ApJ...712L..81K,2008ApJ...686L..49K,2007ApJ...671...40R,2008ApJ...678.1531R}), a scale where inhomogeneities are weak, but still detectable, and we use this distance in our first exploration. For the larger distance we no longer use any scattering law at all, since we wish to test the maximal flux detectable at Earth, when we go through the argument using radio galaxy luminosity functions \citep{2001MNRAS.322..536W}.

\begin{figure}[h]
\centering
\includegraphics[viewport=0cm 0cm 21cm 11cm,clip,scale=0.8]{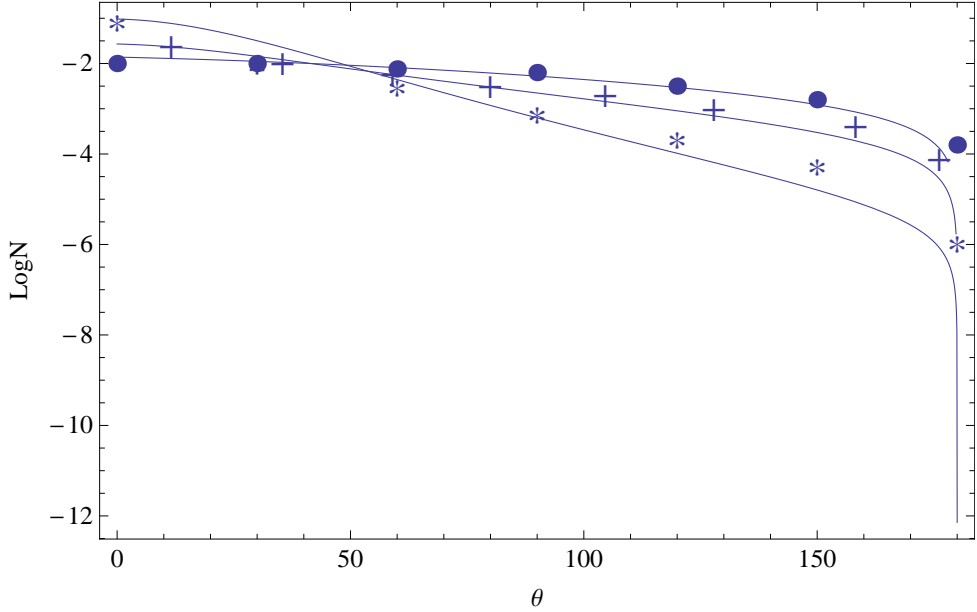}
\caption{Deflection angles distribution from \citet{2008ApJ...682...29D}, the energy range of particles are from 10 EeV to 30 EeV, from 30 EeV to 60 EeV and above 60EeV (points, plus sign and stars).}
\label{DefAngles}
\end{figure}

In the case of particles with high energy, above 10 EeV, we will use the deflection angles from \citet{2008ApJ...682...29D}. Here we will use 
three scattering laws, corresponding to the energy intervals of particles 10 EeV to 30 EeV, between 30 EeV and 60 EeV and beyond 60 EeV (Fig.~\ref{DefAngles}). 
We modeled the scattering laws from the histograms of \citet{2008ApJ...682...29D} as:

$\frac{180^{o}-\theta}{(\theta_{0}^{2}+\theta^{2})^{1/2}}$, where $\theta_{0}=160^{o}$ for the first interval, as
$\frac{180^{o}-\theta}{(\theta_{0}^{2}+\theta^{2})^{1/2}}$, where $\theta_{0}=50^{o}$ for the second and finally as
$\frac{180^{o}-\theta}{(\theta_{0}^{2}+\theta^{2})^{3}}$, where $\theta_{0}=40^{o}$ for the last one.

As the simulations trace only protons we assume here that having heavy nuclei at high energy is akin to using a scattering of low energy protons.

 We assumed that scattering is usually less than 180 degrees,
 implying that scattering runs with $E/Z$,
 \citet{1956pfig.book.....S,2009JCAP...11..009L}, but we need to question this. 

If scattering is low, then for a constant chemical composition across energies any sky-distribution should be reproduced in lower $Z$-elements from higher energy and higher $Z$ elements, as scaled with $E/Z$, allowing beautiful tests of the data.  Of course, if the chemical composition mix is not constant with particle energy as for instance advocated in \citet{2010ApJ...720L.155G}, then any conclusion from data about constancy of scattering with $E/Z$ is moot already.  In this brief paragraph we wish to explore whether it is conceivable, that UHECR particles scatter in the opposite limit, with scattering angles possibly in excess of 180 degrees (or $\pi$ in radians).  Let us consider a filament of the Large Scale Structure (LSS), where the magnetic fields have been estimated to be of order $10^{-7.5}$ Gauss based on non-linear MHD simulations, as quoted in \citet{2010A&A...521A..55C} from the work of \citet{2008Sci...320..909R}.  In such a field the Larmor radius of an Fe particle of $3 \, 10^{20}$ eV is 300 kpc, so less than the width of a LSS filament.  In fact, only for really light elements is the Larmor radius at this energy larger then the width of a filament, of order a few Mpc.  Now, if a radio galaxy such as Cen A were to push such a field into a shell, then the scattering would be reduced as discussed above.  On the other hand, any Larmor radius less than the local scale of the magnetic field structure implies full mirroring at an interface between high and low magnetic field regions, and so no dependence on $E/Z$ anymore.  However, as a consequence the real travel path of any UHECR particle would be drastically increased in turn, and it would be unlikely too observe UHECR particles from almost any plausible source.  Therefore we have two extreme choices:  Either we assume that substructure pushing magnetic fields into thin shells is extreme, and so effective scattering is reduced, or all UHECR particles, which actually make it to us, have traveled outside the high magnetic field regions, so producing an almost blind spot in the direction of any possible nearby source, or effectively enlarging the angular region around it on the sky within which we can expect to observe events to arrive from it.  This second option would reduce the effective flux of events from a nearby source.  Therefore we conclude here, that extreme substructure of the magnetic field may be required, so as to reduce scattering along the path.  Scattering in our magnetic Galactic wind would not introduce any relevant extra path, and might then give the near isotropy observed \citep{2008ApJ...674..258E}.

\section{Interaction of UHECR with the background}

We do not know the exact composition of the ultra high energy cosmic ray particles in the sources. There are several model propositions:

a) As the radio galaxy jets are expanding into the plasma of a small group or a cluster of galaxies, the typically abundances is somewhat less enriched than the ``cosmic abundances" observed near Earth; since Helium is only slightly enriched by galactic evolution, it might be there in near normal abundance. All heavier elements are reduced from normal. We approximate this by 90 percent Hydrogen by number, and 10 percent heavier elements.

b) At the other extreme it has been proposed \citet{2010ApJ...720L.155G,2011arXiv1106.0625B} that the abundances are mostly heavy, as has been suggested for the galactic cosmic rays beyond the knee, near $10^{15} eV$. In this hypothesis the existing galactic cosmic rays are shifted up in energy by the relativistic kick from a relativistic jet boring through the interstellar medium filled with normal cosmic rays, produced by a starburst. In the next tables such radio galaxies are marked with an asterisk.

c) Obviously, other speculative options exist, like pure Hydrogen or pure Iron, but there is no physical model available that would justify such abundances; however, these options are convenient for understanding and discussing propagation.

Next we discuss what happens during propagation with these possibilities:

At the high energies of the cosmic rays, above $3*10^{19} eV$, the light and intermediate nuclei are suppressed by the interactions with CMB photons
and only protons and heavy nuclei are present in the composition of the UHECR. In the following we will work with the hypothesis that the relative abundance at the sources of heavy nuclei is only
$10\%$, and the composition is then dominated by protons, at the level of $90\%$ \citep{2008JCAP...10..033A}.

The effect of infra-red, optical and ultraviolet backgrounds (IR/Opt/UV) photons on the propagated UHECR spectrum of pure protons are almost negligible and the main interaction of protons will be with the cosmic microwave background (CMB) \citep{1966PhRvL..16..748G,1966ZhPmR...4..114Z}.

This is not the case for the heavy nuclei, where the interactions with IR/Opt/UV photons from the background will become significant (e.g. \citet{1999ApJ...512..521S,1976ApJ...205..638P,1969PhRv..180.1264S,1996Rachen,2008JCAP...10..033A}). The interactions experienced by nuclei with photon backgrounds can be summarized as pair production, adiabatic losses and  photo-disintegration processes. The first two result in a decrease of the Lorentz factor of the UHE nucleus, whereas photo-disintegration processes lead to the ejection of one
or several nucleons from the parent nucleus. At different energies ranges we can have different photo-disintegration processes that
become dominant in the total interaction cross section \citep{1976ApJ...205..638P}.

For the present analysis the focus is on the cosmic rays at high and very high energies, in general above 10 EeV produced by sources that are very close to redshift $z = 0$. Using the prescriptions of \citet{2010ApJ...720L.155G} which define a photo-disintegration distance $\Lambda_{dis}$ for any nucleus and energy, on the data of \citet{2008JCAP...10..033A}, we can fit this distance-energy relation for different elements, like Fe, Si, O or He. In this way we can model the nuclear interaction with the background formed by CMB, IR,Opt and UV and use only the part for high energy range of UHECR which is in discussion. This provides a relation between the length till the first interaction of the UHECR produced at the source, with the background, and the energy of the particle.

Using the energy for the production of the particles at the source, we can use the above distance $\Lambda_{dis}$ together with the real distance from us to the source, to get a probability of interaction. As the energy of the particle can drop to $50\%$ after the first interaction, this will basically tell us how many particles will survive from the initial flux at the source to us. Also, for the proton case the same analytic fit procedure can be used for the cross section described in \citep{2001LNP...576.....L}. 

If we were to assume that all RGs associated with starbursts behave as argued for Cen A by \citet{2010ApJ...720L.155G,2011arXiv1106.0625B} there is almost no flux left at high energies, since only a small fraction of the ultra high energy flux is due to either Hydrogen (= protons) or iron. If this were true, then Cen A itself would dominate the overall contribution by 10:1 (see below), and its nuclei would all survive due to the source's proximity. As this possibility has been discussed extensively by these papers already, here we try to generalize basing our arguments on a broader model.

In the following we use the approximation, that for the relatively small distances of the candidate sources the intergalactic magnetic fields do not lengthen the propagation path significantly for Hydrogen; this is obviously a bad approximation for heavy nuclei; for them the propagation path is lengthened so much, that nearly no flux will survive, strengthening the case for our simplified approach.

As we cannot be sure which source can contribute with only protons or heavy nuclei a simplification is made in the sense that the flux of UHECR from each source is estimated as composed from $90\%$ protons and only $10\%$ heavy nuclei. 



\section{Starburst galaxies with GRB}

In starburst galaxies the formation of stars, normal and very massive
stars, is very high e.g.
\citet{1977A&A....54..461B,1985ApJ...291..693K,2009APh....31..138B}.
A small fraction of all massive stars turns into GRB,
which can produce particles with high energy (as suggested qualitatively
by \citet{1994heam.book..217B},
and worked out quantitatively by \citep{1995ApJ...453..883V,1995PhRvL..75..386W}).
As the shock is rapidly slowing down in such explosions, the final
charged particles which are released do not have very high energy.
On the other hand the protons undergoing a proton-photon interaction
and thereby turning into neutrons can be released at very high
energy \citep{1992PhRvL..69.2885P,1998AIPC..428..776R,2004ApJ...604L..29B}.
At a large distance from the source, the neutrons will decay back into
protons (about twenty minutes in the comoving frame, corresponding to
a few Mpc at extreme energies). Under such a hypothesis, most of the
ultra high energy particles may be pure protons, as indeed argued by
the HiRes collaboration \citep{2008APh....30..175T,2010ApJ...713L..64A}.
It has been shown recently, that a revised model of GRB can also produce
heavy nuclei as ultra high energy cosmic rays, see \citet{2008ApJ...677..432W}.
Therefore we emphasize that GRBs offer an option to explain a
pure protons composition, as argued by HiRes, but by no means have to
do this. We also have to note that any conclusion about the proton
fraction is strongly dependent on the interaction model that one
is using.  The GRB might as well mimic the argument proposed in
\citet{2010ApJ...720L.155G}, that Wolf-Rayet stars explode and
give the normal knee cosmic ray spectrum as seeds to further
acceleration by a relativistic shock driven by a central
super-massive black hole. Therefore GRBs might even give a similar fit to the Pierre Auger Observatory data as suggested in \citet{2010ApJ...720L.155G}, provided GRB make a considerable contribution to cosmic rays
\citep{2005ApJ...628L..21D}.

We consider in this section the possibility that GRB may be the
majority of sources of the ultra high energy cosmic rays.
Starburst galaxies are best selected at 60 microns where high
luminosity in the far infra-red implies a high star formation
rate, \citet{1977A&A....54..461B,1985ApJ...291..693K,1981JRASC..75R.249K,1996ARA&A..34..749S},
and a high supernova rate. It follows that we have a high GRB rate
in such galaxies.

\subsection{Sample of sources}

 We give a complete sample in Table ~\ref{table60micron} of 31 candidate
 sources, up to redshift 0.0125, above 50 Jy, all far-infrared selected
 galaxies including starburst galaxies. These selection parameters are chosen in order to have a comparable number between the GRB and Radio galaxies sources. Further selections at higher redshift can be made to explore the contribution of distant sources, this approach will be investigated in the future.
We give the name, distance, flux
 density at 60 $\mu$, flux density at 5 GHz, FIR/radio ratio, and
 color B-V.  The FIR/radio ratio shows a relatively small range due to
 the well-established correlation between these two flux densities
 \citep{1985A&A...147L...6D,1985ApJ...291..693K,1996A&A...306..677L,1996A&A...314..745L,1989ESASP.290..531V,1994A&A...282...19X,1994A&A...285...19X,2010HiA....15..417T}.
 This is not adhered to by the radio galaxy NGC5128, which has huge radio
 lobes, powered by a central super-massive black hole.
 The values of the color index B-V shows that these galaxies have a
 young stellar population dominating the emission (B-V $<$ 0.5), and
 few are extremely reddened \citet{1977A&A....54..461B} or show
 dominance of an old stellar population (B-V $\simeq$ 1) with M82
 highly reddened, and M31 just old.  These galaxies are expected
 to show strong star formation, and so a relatively high rate of GRB events. We have included the radio galaxy NGC5128 = Cen A,
 since its center also contains a starburst.

It is noteworthy that here the distances are very small, typically less than 7 Mpc, so interaction in the bath of the microwave background for those close sources are weak for Hydrogen. 


\begin{table}
\begin{center}
\small
\caption{Properties of the selection at 60$\mu$m, redshift $z\leq 0.0125$, flux density brighter than 50 Jy, starburst selected, sample of 32 candidate sources}
\begin{tabular}{|c|c|c|c|c|c|}
\hline\hline
Name&Distance&Flux density&Flux density&FIR/Radio&B-V\\
&Mpc&at 60$\mu$m(Jy)&at 5GHz(Jy)&ratio&mag\\
\hline
MESSIER 031&0.72&536.18&1.79&299.5&0.68\\
MESSIER 033&0.78&419.65&1.3&322.8&0.47\\
NGC 1569&1.45&54.36&0.16&350.7&-\\
NGC 6946&5.64&129.78&0.34&377.2&0.4\\
NGC 0055&1.34&77&0.16&481.25&0.54\\
SMC&0.06&745&-&&0.41\\
MESSIER 082&5.68&1417.1&3.91&362.4&0.79\\
NGC 0253&3.22&967.81&2.08&465.2&-\\
LMC&0.05&82917&-&&0.44\\
MESSIER 094&4.4&71.54&0.25&286.16&0.72\\
Circinus Galaxy&3.13&248.7&0.61&407.7&-\\
MESSIER 051a&8.33&97.42&0.36&270.6&0.53\\
NGC 0891&10.23&66.46&0.24&273.4&0.55\\
NGC 5128&3.42&213.29&681&0.31&-\\
NGC 2903&6.74&60.54&0.18&336.3&0.55\\
NGC 4945&4.85&625.46&2.8&223.34&-\\
MESSIER 051&7.89&108.68&0.53&207&-\\
NGC 4631&6.93&85.4&0.34&250.4&0.55\\
MESSIER 066&6.12&66.31&0.18&368.3&0.6\\
NGC 3628&7.9&54.8&0.2&274&0.54\\
NGC 0660&11.6&65.52&0.16&420&0.71\\
NGC 2146&12.2&146.69&0.43&338.7&0.63\\
NGC 1808&13.6&105.55&0.21&502.6&0.74\\
NGC 3079&15.2&50.67&0.33&153.55&0.67\\
MESSIER 077&15.5&196.3&1.89&103.8&0.7\\
NGC 1097&17.4&53.35&0.15&355.6&0.68\\
NGC 7582&21.5&52.2&0.11&474.5&0.6\\
NGC 7552&22&77.37&0.14&552.6&0.63\\
NGC 1365&22.4&94.31&0.21&449.01&0.59\\
NGC 3256&38.4&102.63&0.24&427.6&0.44\\
ESO 173- G 015&39.9&81.44&0.15&542.9&-\\
NGC 3690&42.7&113.05&0.36&312.2&-\\
\hline
\end{tabular}
\label{table60micron}
\end{center}
\end{table}


We start with an initial number of particles at the sources of 300; this is motivated by the observed flux in the chosen energy range
as we will compare the our numbers with the observed one at Earth. As the number of observed particles will increase in the future, 
we can easily change this initial number and check again.
 
Since the flux of ultra high energy particles from
 GRB is extremely unsteady, we run a dual Monte-Carlo, one step for the
 probability, that we are within the duty cycle to actually observe
 particles (\citet{2004ApJ...604L..29B}), and a second one to give the
 flux of the GRB for each source that is in the duty cycle. The first one is based on the luminosity at 60$\mu$m
and is basically a random choice weighted by the luminosity, equivalent with a Monte-Carlo sampling method from the distribution of the weights.
 This randomly distributes the initial 300 particles over the list of sources 
according to the weights. If a source has a non zero number then it will be in the duty cycle.
The second Monte-Carlo is using the flux at 60$\mu$m as weights to randomly distributes the initial flux of particles to the sources which are active.

 Including the two Monte-Carlo steps for the angular scattering, these four Monte-Carlo steps have important consequences. The most
 important effect is that the duty cycle gives a ``yes" or ``no" answer
 for any galaxy, whether we are in ``luck" to observe a GRB now.
 This can be understood as the ratio of the time scale a GRB is
 visible in UHECR, as broadened by magnetic scattering of the
 particles, to the typical time scale between GRB, which in our model
 scales inverse with the 60 microns luminosity. Taking as an example
 LMC, this implies that we have almost no chance to ever observe a
 GRB there.  This first step implies, due to the flat flux density
 distribution of the galaxies in our sample, that different
 runs can give very different sky distributions. Given the first
 Monte-Carlo step, the second then just gives the rate of particles
 produced by the sources.

Regarding the composition of particles at the source we follow the work of \citet{1993A&A...271..649B} and \citet{2011arXiv1106.0625B} in which the flux of the nuclei is divided in 
element group, like He, CNO, Ne-S, Cl-Mn, and Fe and we take the ratios of this numbers from the energy spectrum, so for instance, above $8\times 10^{19}$ Fe dominates in all the ratios. 

\subsection{The events from GRB on the sky and the energy spectrum}

Having the sources that will accelerate particles and the numbers of cosmic rays per source, we can start the simulation of their journey towards Earth.
First we start with a $E^{-2}$ distribution of energy at the source for the initial flux of cosmic rays, between 30 EeV and 200 EeV. We associate our 
particles the energies according to this distribution by sampling the above distribution of energies. Following the prescription from the section 3, we simulate the interaction with the CMB and IR/Opt/UV backgrounds
to get the probability of interaction of each particle. Based on this we determined the number of particles which will have no interaction and will arrive at Earth with the same energy as at
the source. In the following we select only these particles and because the rest will significantly lose their energy, we leave the ones which have a $100\%$ probability of interaction or even multiple interaction with the background for future modeling.

The next step is to simulate the deflections in the intergalactic magnetic fields of the surviving particles. We separate the particles by their energies and use the scattering laws
 described before for each energy interval. This procedure will randomly associate particles in a certain energy bin with a population of deflection angles. The final step is to 
distribute the events around each original source using the scattering angles obtained in the previous step in uniform distributed rings around each source.

\begin{figure}[h]
\centering
\includegraphics[viewport=0cm 0cm 21cm 11cm,clip,scale=0.8]{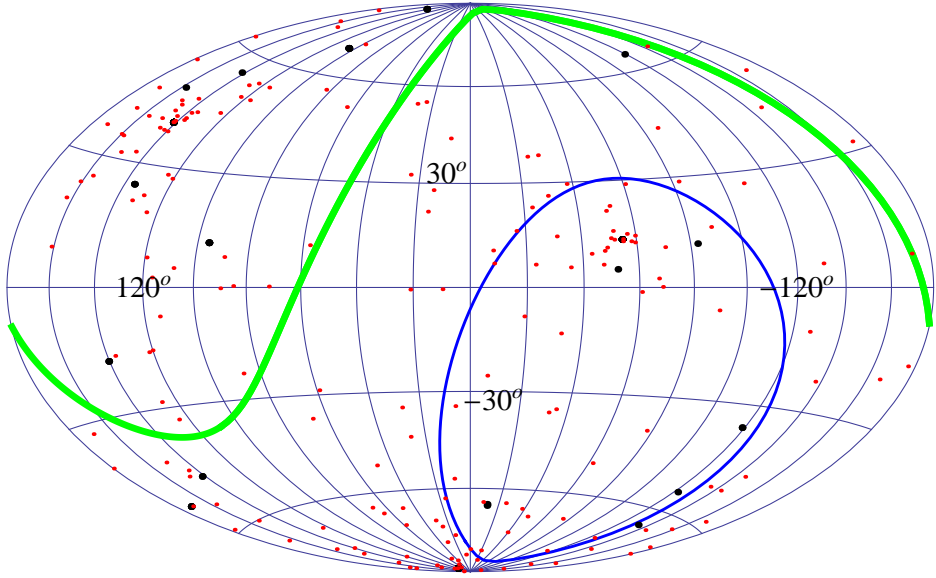}
\caption{Aitoff projection in galactic coordinate of the Monte Carlo scattered 300 events (red dots) coming from a selected population of sources of GRB (black dots). Above the thick line is the area from the sky not seen by Pierre Auger Observatory (declinations above than 24.8 degrees) and the thin line surrounds the area from the sky not seen by the HiRes experiment (declinations less then -32 degrees). The main contributors are M82, NGC 0253 and NGC 4945 in this order.}
\label{GRBscattering}
\end{figure}

We present the final sky distribution which shows the sources that
 are inside the duty cycle and are accelerating particles in
 Fig.~\ref{GRBscattering}. The coordinate system used here is
 with Galactic plane running horizontally, and with the Galactic Center
 at the center. Generated simulations can look rather
 different, due to the duty cycle suppression. As expected, there is a weak clustering of events
 around the most productive sources.

 The figures mainly show that slightly more particles are expected in
 the North than in the South (see Table~\ref{ContributionTable}).
 In fact, this specific class of sources is the only one, which allows
 more events to be detected in the North than in the South, as long as
 the scattering is not giving complete homogeneity.

\begin{figure}[htpb]
\centering
\includegraphics[viewport=0cm 0cm 21cm 11cm,clip,scale=0.8]{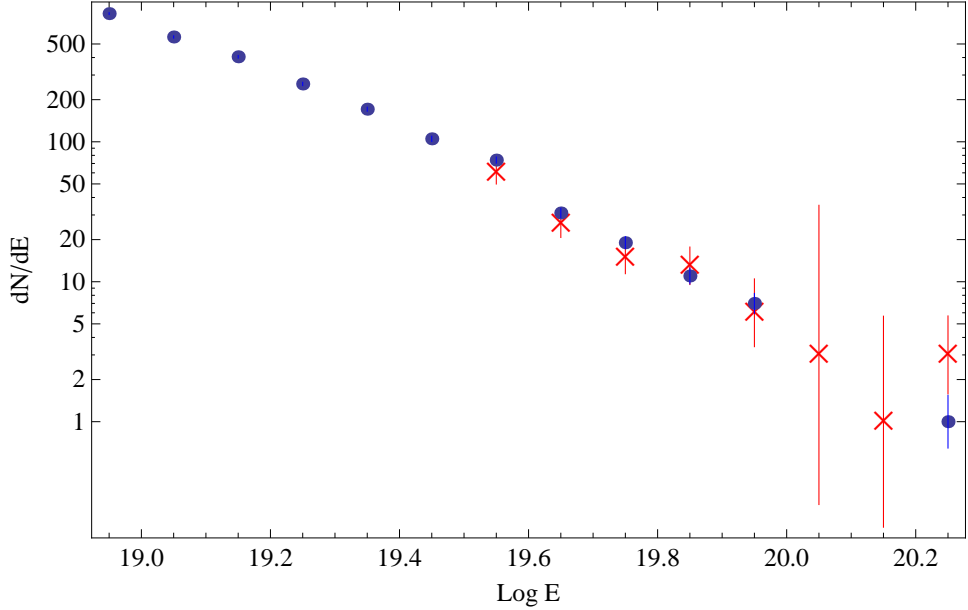}
\caption{Energy spectrum of the Pierre Auger Observatory in blue (The Pierre Auger Coll., Phys. Rev. Lett., vol. 101, 6 (2008)) and the corresponding energy spectrum coming from a selected GRB list of sources represented by a star symbol.}
\label{EnergySpectrumGRB}
\end{figure}

In order to obtain the energy spectrum we select from the arriving particles at the Earth the ones that can be observed from Pierre Auger Observatory (declinations below than 24.8 degrees).
We compare the number of cosmic rays per energy bin between the Pierre Auger Observatory data \citet{2008PhRvL.101f1101A} and our simulated ones (Fig.~\ref{EnergySpectrumGRB}). This is because it is quite difficult to compare the flux of cosmic rays as a function of energy for the observed and simulated data. As the flux is given in terms of exposure and this is turn is influence by many factors, like the efficiencies of the various steps of the data analysis, the evolution of the detector in the construction phase (see \citet{2010PhLB..685..239A} for further details), the comparison between the flux of real data and the simulated one will be made in a future article. For our simulated data we added at the Poisson errors the standard deviation at each point that resulted from individual sets of 10 simulations. Future effort will be made to increase the number of simulations to significant statistical level, $>10^{5}$.

\subsection{The cutoff spectrum?}

 Since we do see a turn-off in the spectrum, \citet{2008PhRvL.101f1101A}, we need to ask, whether
 this turn-off could be due conditions in the source, or whether
 it could be due to GZK interactions with the microwave background.
 To have such interactions in the source, and still be visible at
 Earth, we would need to have most of the sources nearby, so that such
 a spectral feature is not washed out. Also we need a photon density
 in the source far above the microwave background, that the survival
 time of the source is short enough to exceed any effect from the
 CMB.  This implies a photon density in the source, by the ratio of
 the interaction time in the microwave background, about
 $30 \times 10^{7}$ years to the lifetime of a GRB, about 1 second
 in the observer frame, so more than $10^{16}$ times.
 This appears easily possible in a GRB.  When the GRB becomes visible
 in electromagnetic emission, the source becomes optically thin to
 photon-photon absorption in the comoving frame.  Since the proton-photon
 cross-section is smaller than the Thompson cross section, the source
 will become optically thin for protons earlier than for the photons,
 \citet{1998AIPC..428..776R}, and this will result in a slight advance
 from the resulting neutrinos.  In the comoving frame the required
 nucleus energy is lower, and the implied photon energy is as large as
 to reach the threshold for pion production. So the observed X-ray photons
 may give rise to a turn-off in the observed spectrum near 40 EeV.
 When the source becomes progressively optically thin, we accumulate
 particles from above the threshold.  This may qualitatively explain
 the steeper slope beyond about 40 EeV suggested by the data.

\section{Radio galaxies}

 Radio galaxies have been under prime suspicion to be the sources of
 UHECR for many years, as first argued by \citet{1963SvA.....7..357G}. They can
 accelerate not just the electrons which give rise to synchrotron
 emission in the radio range, but also protons and other nuclei.
 Quantitative arguments based on observations were made some time ago
 (\citet{1987ApJ...322..643B}), and show that radio galaxies such as M87
 produce particles near $10^{21}$ eV and this is required to explain
 a very common cutoff seen in the optical non-thermal continuum spectra
 near normally $3 \times 10^{14} \, {\rm Hz}$. We note that in a
 photon-strong environment this cutoff could be shifted to higher
 frequencies by a factor of order 5.  If the emission region is
 boosted by relativistic motion, the frequency could be yet higher.

 In order to test the idea that RGs are source candidates
 we have developed the jet-disk symbiosis concept e.g., papers by
 \citep{1995A&A...293..665F,1995A&A...298..375F,1999A&A...344L..37P,2001A&A...372L..25M,2002A&A...383..854Y,2008A&A...477....1M}.
 This allows us to estimate the flux in ultra high energy cosmic ray particles.

\subsection{Complete samples}

 Here we use a complete sample of steep spectrum radio sources,
 selected at 5 GHz  \citep{1971A&A....11..171W,1978AJ.....83..451P,1981A&AS...45..367K,1981AJ.....86..854K,1984AJ.....89..323G,1986A&A...168...17E}.
 In Table~\ref{tableSteepSources} we give a complete sample of all
 extragalactic steep spectrum sources down to a flux density limit
 of $0.5$ Jy. We exclude those which are dominated by a starburst,
 already known from their far-infrared/radio flux density ratio,
 and impose the condition $z\leq 0.025$.
 This is a complete sample of steep spectrum radio sources, 
 selected at 5 GHz, excluding normal galaxies, and also redshift
 limited.  As a control we give the FIR/radio ratio, which in no
 case comes close to the standard value for normal and starburst galaxies (several
 hundred, see Table~\ref{table60micron}).  We again give name,
 distance, black hole mass, compact radio flux density if known,
 color B-V, and as noted FIR/radio ratio.  The steep spectrum
 selection uses $\nu^{-0.5}$ as the cut-off criterion
 \citep{1984AJ.....89..323G}. We show a complete sample of flat
 spectrum radio sources in Table~\ref{tableneutrino}, selected with a spectrum flatter than the cut-off.  Radio galaxies
 are usually early Hubble type galaxies, and have a color B-V which
 is red (i.e. near 1.0).  Exceptions can be starbursts induced by
 a merger of an elliptical with a spiral, but then reddening has
 modified the color again, so as to render it red, as appears to be
 the case with NGC5128 (Cen A).  NGC4651 is the only case with a
 clearly blue color, and it also has a relatively large FIR/radio ratio
 as compared to other radio galaxies, these have a modest starburst,
 NGC5793, IC5063, and NGC4651; NGC5128 (= Cen A) has a quite modest
 starburst as measured by its FIR/radio ratio.

\begin{table}[h!]
\begin{center}
\small
\caption{Properties of the selection in passband 6cm (5 GHz), redshift
$z\leq 0.025$ flux density brighter than 0.5 Jy, steep spectrum and no
starburst, sample of 29 candidate sources}
\begin{tabular}{|c|c|c|c|c|c|}
\hline\hline
Name&$D_{L}$&M$_{BH}$&$S_{core}$&B-V&FIR/Radio\\
&Mpc&$10^{8}$M$_{\odot}$&mJy&mag&ratio\\
\hline
NGC 5128&3.4&0.55&8600&0.88&0.313\\
NGC 4651&18.3&0.4&0&0.51&-\\
MESSIER 084&16&15&168.7&0.94&0.174\\
MESSIER 087&16&31&3097.1&0.93&0.005\\
NGC 1399&15.9&5.1&0&0.95&0.044\\
NGC 1316&22.6&5.1&0&0&0.063\\
NGC 2663&32.5&8.23&0&0&0.084\\
NGC 4261&16.5&5.2&0&0.97&0.019\\
NGC 4696&44.4&11.1&0&0&0.076\\
NGC 3801&50&1.95&0&0.9&0.298\\
IC 5063&44.9&2.32&0&0.93&-\\
NGC 5090&50.4&8.95&0&0&0.099\\
NGC 5793&50.8&0.3&0&0.79&-\\
IC 4296&54.9&2.53&0&0.95&0.079\\
NGC 0193&55.5&2&0&0.98&0.76\\
VV 201&66.2&1&0&0&0.052\\
UGC 11294&63.6&2.9&0&0&0.326\\
NGC 1167&65.2&5.42&5.9&0&0.133\\
CGCG 114-025&67.4&2.19&0&0&0.014\\
NGC 0383&65.8&6.67&0&0&0.207\\
ARP 308&69.7&1&89.7&0&0.088\\
ESO 137- G 006&76.2&15.1&130&0&0.021\\
NGC 7075&72.7&2.5&0&0.97&0.196\\
UGC 02783&82.6&4.2&0&0.99&0.924\\
WEIN 045&84.6&4.59&0&0&0.92\\
UGC 01841&84.4&1&0&0&0.035\\
NGC 3862&93.7&6.74&171&0.94&0.106\\
NGC 1128&92.2&2&0&1.02&0.056\\
NGC 5532&104.8&10.8&0&0&0.022\\
\hline
\end{tabular}
\label{tableSteepSources}
\end{center}
\end{table}

\subsection{Particle energy and particle flux predictions}

 The jet-disk hypothesis has successfully been able to account for
 observations, such as spectra across the entire electromagnetic range,
 temporal variability, and spatial structure.  It was mainly tested on
 low power sources, because they are common, and also more challenging
 \citep{1995A&A...293..665F,1995A&A...298..375F,1996A&A...308..321F,1996A&A...316...43D,1999A&A...342...49F,2000A&A...362..113F,2001A&A...372L..25M}.

 The main indefinite parameter in the jet-disk symbiosis picture is
 the anchoring of the magnetic field at the base of the jet.
 Spin-down powered jets emanating from near super-massive black holes,
 \citet{1977MNRAS.179..433B,1979ApJ...232...34B,1999MNRAS.307..491B}
 are one possibility to do this.  The jet power scales with the total
 radio luminosity, \citet{1997ApJ...477..560E}, suggesting a range of
 possible power-law exponents, from 0.7 to 1.0, Table~\ref{tableJetPower}.
 For the previously defined complete sample we check various simple
 properties, and then, based on these, we modeled them in
 Table~\ref{tableSteepSources}. We give first the name again, then the
 Eddington luminosity using the BH mass, derived from observations or
 from the galaxy-BH correlations, \citet{2010A&A...521A..55C}, then the
 ratio $L_{jet}/L_{Edd}$ using four different approaches, first using the
 correlation between radio-power and jet power with an exponent of
 $\alpha = 1.0$, then with an exponent of 0.7, then using an empirical
 formula due to \citet{1988A&A...199...73G}, and finally from observed
 compact radio emission, where available. In the last column we also
 check on the ratio $L_{X}/L_{jet}$.  In the case of M87, for which
 good data on the compact emission is available, we used those data
 exclusively.  The main result is that the four different approaches
 give similar results, to within better than usually an order of
 magnitude total spread.  The last column shows that the X-ray emission
 is a small part of the energy budget, demonstrating that almost all
 these sources are in low accretion limit, as even the most extreme
 source is below $0.1 L_{Edd}$, with most of the sources far below
 this value.  Typically RGs seem to be in a low accretion
 limit, consistent with the assumptions made in \citet{2010arXiv1012.0204B}.

 If we identify the jet power as an upper limit to the Poynting flux,
 and use the relationship between Poynting flux and maximal particle
 energy (\citet{1976Natur.262..649L}), we obtain an expression for the
 maximum particle energy.  Furthermore we can assume that the cosmic
 ray flux is a fraction of the total jet power, and so we obtain a
 simple proportionality.

 We call in the following the mass of the black hole $M_{BH}$, the
 observed compact radio flux density $S_{core}$ or extended total
 flux density at 5 GHz $S_{5GHz}^{extended}$, the luminosity distance
 $D_L$ to the radio galaxy, the  ultra high energy cosmic ray luminosity
 at the Eddington limit, the maximum particle energy $E_{max}$, and
 the maximum cosmic ray flux $F_{CR}$.

 The estimated flux it is a maximum flux and this is
 especially true, since the Poynting flux is only a minimum energy
 flow along the jet, and strongly constrains the maximum particle energy.

 Interestingly, the mass of the black hole does not even enter here
 due to the simplicity of the Poynting flux argument. Of course, in its theoretical derivation the Poynting flux is directly proportional to the black hole mass, but here we use observations to obtain a limit on the Poynting flux, and so indirectly infer also a limit on the black hole mass. Following this
 below we might have to multiply the maximum particle energy by a
 factor of order 10 to simulate the seeding with heavier nuclei from
 a weak starburst, indicated by a relatively large FIR/radio ratio as
 given in Table~\ref{tableSteepSources}. This ratio is still below
 that of a pure starburst, for which it is of order 300.  Our numbers,
 given for the maximum energy, do not include this extra factor and we
 marked the corresponding sources that can have seeding in
 Table~\ref{tablepredictions} with an asterisk.

 In Table~\ref{tablepredictions} we show the derived estimates for the
 maximum particle energy, and for the maximum particle fluxes.
 The marked galaxies may have sufficient starburst power to allow seeding
 with heavy nuclei; however, the numbers derived are all assuming
 protons.  We again give first the name, then three version of
 maximum particle energy $E_{max}$, and two version of maximal
 particle flux $F_{max}$, all relative to M87, the prototype radio galaxy
 in our cosmic neighborhood.  The first two version of $E_{max}$
 for the assumption of spin-down powered jets, using $\alpha = 1$ and 0.7,
 then a column assuming that accretion power drives the jets, and in
 the columns for maximal fluxes we use spin-down with $\alpha = 1$,
 and accretion power.  We note that accretion power does give quite
 different numbers in these approaches from spin-down power, but the
 rank of the first four radio galaxies in predicted flux is
 NGC5128 = Cen A, NGC4486 = M87 = Vir A, NGC1316 = For A, and NGC4261
 for spin-down power, and for accretion power exactly the same.
 The first three were already identified by \citet{1963SvA.....7..357G}
 as the most likely sources to produce an abundant flux of extra-galactic cosmic rays.
 The maximum particle energies achieved in the two approaches,
 spin-down versus accretion powered give quite different results:
 spin-down gives systematically higher energies, whereas accretion
 gives relatively low energies for most sources, insufficient to
 explain the data. Also, we note that using heavy nuclei of course has
 the potential to increase considerably the maximum energy.

 Accretion powered jets are the other alternative, which works well for
 relatively high current accretion rates
 \citep{1995A&A...293..665F,1995A&A...298..375F,2004TascauThesis}.

 For distances $<$ 50 Mpc usually NGC5128 = Cen A, possibly NGC1316 = For A,
 and a group around M87 = Vir A dominate the predicted UHECR flux
 \citep{1963SvA.....7..357G}.  The first five in flux density of the
 extended flux are ESO137-G006, NGC1316, NGC4261, NGC4486=M87, and NGC5128.

\begin{table}
\begin{center}
\small
\caption{Jet powers}
\begin{tabular}{|c|c|c|c|c|c|c|}
\hline\hline
Name&$L_{Edd.}$&$L_{jet}/L_{Edd.}$&$L_{jet}/L_{Edd.}$&$L_{jet}/L_{Edd.}$&$L_{jet}/L_{Edd.}$&$L_{X}/L_{jet}$\\
&$10^{47}erg/s$&$\alpha=1$&$\alpha=0.7$&Giovannini&from compact&$10^{-3}$\\
(1)&(2)&(3)&(4)&(5)&(6)&(7)\\
\hline
NGC 5128&0.077&0.02342&0.03067&0.03587&0.01452&1.205\\
NGC 4651&0.056&0.00095&0.00357&0.02423&-&0.417\\
MESSIER 084&2.1&0.00008&0.00021&0.00061&0.0003&0.044\\
MESSIER 087&4.34&0.00102&0.00102&0.00102&0.00102&0.085\\
NGC 1399&0.714&0.00006&0.00025&0.00083&-&80.905\\
NGC 1316&0.714&0.00403&0.00458&0.00494&-&0.210\\
NGC 2663&1.152&0.0002&0.00048&0.00114&-&-\\
NGC 4261&0.728&0.00224&0.00302&0.00377&-&0.225\\
NGC 4696&1.554&0.00029&0.00058&0.00141&-&0.053\\
NGC 3801&0.273&0.00118&0.00259&0.00586&-&-\\
IC 5063&0.325&0.00074&0.00178&0.00622&-&34.386\\
NGC 5090&1.253&0.00056&0.00097&0.00211&-&-\\
NGC 5793&0.042&0.007&0.01574&0.04117&-&-\\
IC 4296&0.354&0.00343&0.00506&0.00549&-&4.916\\
NGC 0193&0.28&0.00162&0.00321&0.00663&-&-\\
VV 201&0.14&0.02041&0.02328&0.02487&-&-\\
UGC 11294&0.406&0.00156&0.00279&0.00551&-&-\\
NGC 1167&0.759&0.00114&0.00186&0.00293&0.00058&2.071\\
CGCG 114-025&0.307&0.00866&0.01009&0.01147&-&0.607\\
NGC 0383&0.934&0.0022&0.00277&0.00307&-&0.099\\
ARP 308&0.14&0.01469&0.01849&0.02508&0.02122&1.939\\
ESO 137- G 006&2.114&0.00451&0.00358&0.00283&0.00203&-\\
NGC 7075&0.35&0.00174&0.00316&0.00526&-&-\\
UGC 02783&0.588&0.00142&0.00234&0.00341&-&-\\
WEIN 045&0.643&0.00545&0.00585&0.00758&-&-\\
UGC 01841&0.14&0.04286&0.03912&0.03439&-&0.560\\
NGC 3862&0.944&0.00419&0.00434&0.00488&0.00718&8.185\\
NGC 1128&0.28&0.01609&0.016&0.01268&-&-\\
NGC 5532&1.512&0.00279&0.00283&0.00258&-&0.147\\
\hline
\end{tabular}
\label{tableJetPower}
\end{center}
\end{table}

\begin{table}
\begin{center}
\small
\caption{UHECR predictions}
\begin{tabular}{|c|c|c|c|c|c|c|}
\hline\hline
Name&$S^{extended}_{5GHz}$&$E_{max}^{\Downarrow}/E^{M87}_{max}$&$E_{max}^{\Downarrow*}/E^{M87}_{max}$&$E_{max}/E^{M87}_{max}$&$F_{max}^{\Downarrow}/F^{M87}_{max}$&$F_{max}/F^{M87}_{max}$\\
&mJy&spin down&spin down&acc. dom.&spin down&acc. dom.\\
(1)&(2)&(3)($\alpha=1$)&(4)($\alpha=0.7$)&(5)&(6)($\alpha=1$)&(7)\\
\hline
NGC 5128*&681000&0.638&0.730&0.013&8.917&12.029\\
NGC 4651*&700&0.110&0.213&0.003&0.009&0.040\\
MESSIER 084&2880&0.194&0.318&0.162&0.038&0.112\\
MESSIER 087&76370&1.000&1.000&1.000&1.000&1.000\\
NGC 1399&720&0.101&0.201&0.036&0.009&0.043\\
NGC 1316&49000&0.806&0.860&0.142&0.642&0.741\\
NGC 2663&950&0.227&0.354&0.099&0.012&0.033\\
NGC 4261&8320&0.606&0.705&0.120&0.109&0.152\\
NGC 4696&1320&0.320&0.450&0.167&0.017&0.037\\
NGC 3801*&570&0.270&0.400&0.026&0.007&0.018\\
IC 5063*&530&0.234&0.362&0.028&0.007&0.018\\
NGC 5090&1710&0.397&0.523&0.156&0.022&0.041\\
NGC 5793*&508&0.259&0.388&0.004&0.007&0.016\\
IC 4296&1780&0.524&0.636&0.053&0.023&0.036\\
NGC 0193*&650&0.320&0.450&0.030&0.009&0.018\\
VV 201&2880&0.803&0.858&0.028&0.038&0.044\\
UGC 11294*&690&0.378&0.506&0.049&0.009&0.017\\
NGC 1167&901&0.443&0.565&0.102&0.012&0.020\\
CGCG 114-025&2580&0.774&0.836&0.060&0.034&0.040\\
NGC 0383*&2100&0.682&0.765&0.167&0.027&0.035\\
ARP 308&1870&0.682&0.765&0.025&0.024&0.032\\
ESO 137- G 006&7250&1.467&1.308&0.629&0.095&0.074\\
NGC 7075&510&0.371&0.500&0.042&0.007&0.013\\
UGC 02783*&541&0.435&0.558&0.078&0.007&0.012\\
WEIN 045*&2160&0.889&0.921&0.137&0.028&0.031\\
UGC 01841&3720&1.164&1.112&0.036&0.049&0.044\\
NGC 3862&1990&0.945&0.961&0.209&0.026&0.027\\
NGC 1128&2340&1.009&1.006&0.065&0.031&0.030\\
NGC 5532&1698&0.977&0.984&0.343&0.022&0.023\\
\hline
\end{tabular}
\label{tablepredictions}
\end{center}
\end{table}

The maximal energy for M87 in the model that Hydrogen is the dominant component is about 100 EeV, using the relationship connecting the Poynting flux limit with the maximal particle energy containable (Lovelace 1976), allowing for some variability perhaps 200 EeV. In the accretion dominance model most of the other maximal article energies are lower, often much lower; in the spin-down dominance model almost all the other maximal energies are slightly lower. In both cases the sources with strongest predicted flux are at distances less than about 25 Mpc, so a quarter of the maximal distance in the sample. So any modification of the maximal particle energy for the weak sources becomes irrelevant in the sum. This is important since only very few sources make up the dominant summed flux in the case of RGs.

 These two approaches may account for the huge range in radio to
 optical flux ratios from active galactic nuclei (\citet{1980A&A....88L..12S}),
 and the ubiquity of low flux densities of compact radio emission from
 basically all early Hubble type galaxies e.g. \citep{1984A&A...130L..13P}.
 As soon as the accretion rate drops below the critical level,
 spin-down takes over from accretion as the powering mode. As pointed out
 by Blandford \citep{1977MNRAS.179..433B} the decay time of the spin-down
 powered activity is very long, see also \citet{1999ApJ...523L...7A},
 and may allow to understand the appearance of ``inverse evolution" for
 flat spectrum radio sources. The activity per comoving starburst galaxies
 and also the central activity in galaxies, peaked in the redshift range
 between 1.5 to 2 and since then the activity decreased by about a factor
 of 30, and at present we have a population of early Hubble type galaxies
 which had their prime activity long time ago. If all central black holes
 stay active - as observations suggest, albeit at a very low level - then
 a subset of the population of these black holes will aim their jet at Earth,
 and so give rise to a weak, but dominant flat radio spectrum.  One
 consequence of such a speculation is that most central super-massive black
 holes should be close to the maximal spin.

 All these weakly active galactic nuclei will also accelerate particles
 to high energy, but will have an extremely low flux.  Such particles would
 have to be injected from the interstellar medium of the early Hubble-type
 host galaxy, and their composition is mostly protons, with a trace of
 Helium.  Also their maximum energy will be relatively low.

\begin{figure}[htpb]
\centering
\includegraphics[viewport=0cm 0cm 21cm 11cm,clip,scale=0.8]{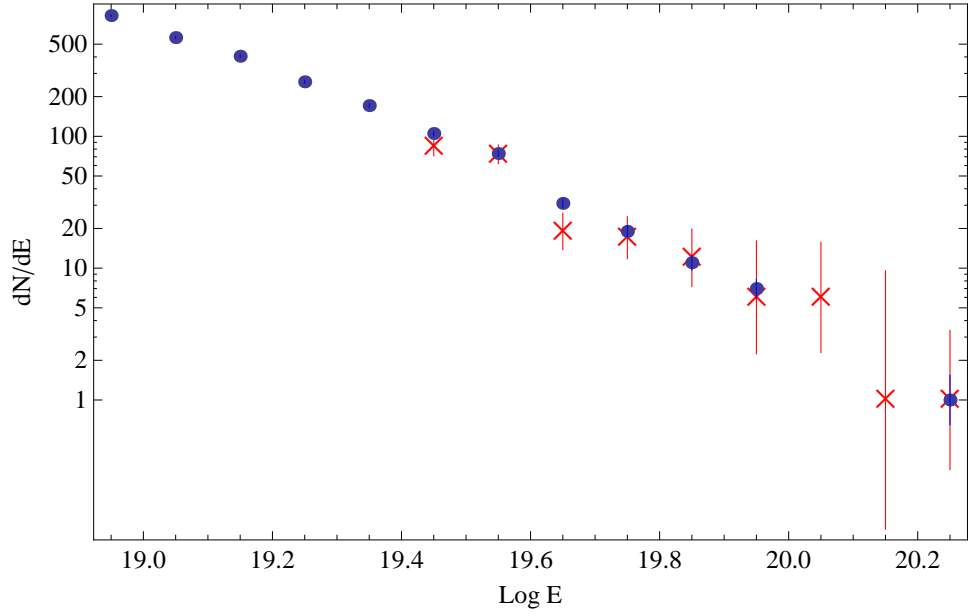}
\caption{Energy spectrum of the Pierre Auger Observatory in blue (The Pierre Auger Coll., Phys. Rev. Lett., vol. 101, 6 (2008)) and the corresponding energy spectrum coming from a selected Radio galaxies list of sources represented by a star symbol. Here the acceleration mechanism is accretion dominated.}
\label{EnergySpectrumAGNAccretion}
\end{figure}

\begin{figure}[htpb]
\centering
\includegraphics[viewport=0cm 0cm 21cm 11cm,clip,scale=0.8]{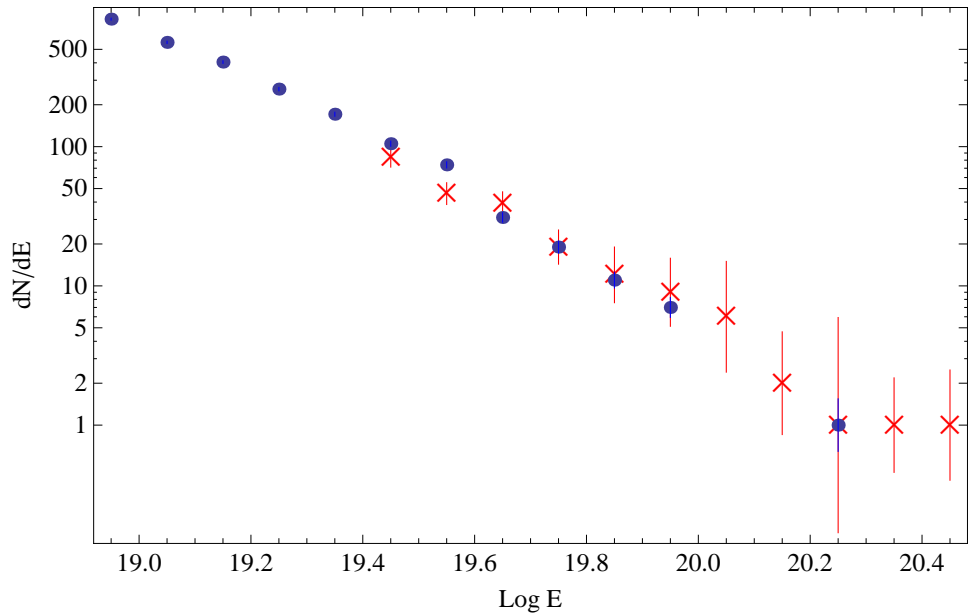}
\caption{Energy spectrum of the Pierre Auger Observatory in blue (The Pierre Auger Coll., Phys. Rev. Lett., vol. 101, 6 (2008)) and the corresponding energy spectrum coming from a selected Radio galaxies list of sources represented by a star symbol. Here the acceleration mechanism is spin-down.}
\label{EnergySpectrumAGNSpinDown}
\end{figure}

We follow the same approach as in the case of GRB regarding the simulations with the difference that the maximal energy of the particles from each source is provided by the Table~\ref{tablepredictions} with the reference energy of the M87 set at the level of $10^{21}$ eV. Also the distribution of the energy follows a $E^{-3}$ law. Regarding the composition at the sources we follow the same approach like in the case of GRB, with the note that here, because the energy levels are different for each source, we get a increase in the mix of heavy nuclei. The resulting energy spectrum for the two acceleration mechanisms is presented in Fig.~\ref{EnergySpectrumAGNAccretion} and Fig.~\ref{EnergySpectrumAGNSpinDown}.

\subsection{Sky distributions}

With a simple prescription for the angle scattering as described above we can turn a source list with predicted cosmic ray fluxes into probable sky distributions by generating sky maps of different numbers of events coming from the proposed sources. As before we selected 300 events to give a illustrative picture of the sky on different assumptions and to be at the level of observed events in the selected high energy range.

\begin{figure}[htpb]
\centering
\includegraphics[viewport=0cm 0cm 21cm 11cm,clip,scale=0.8]{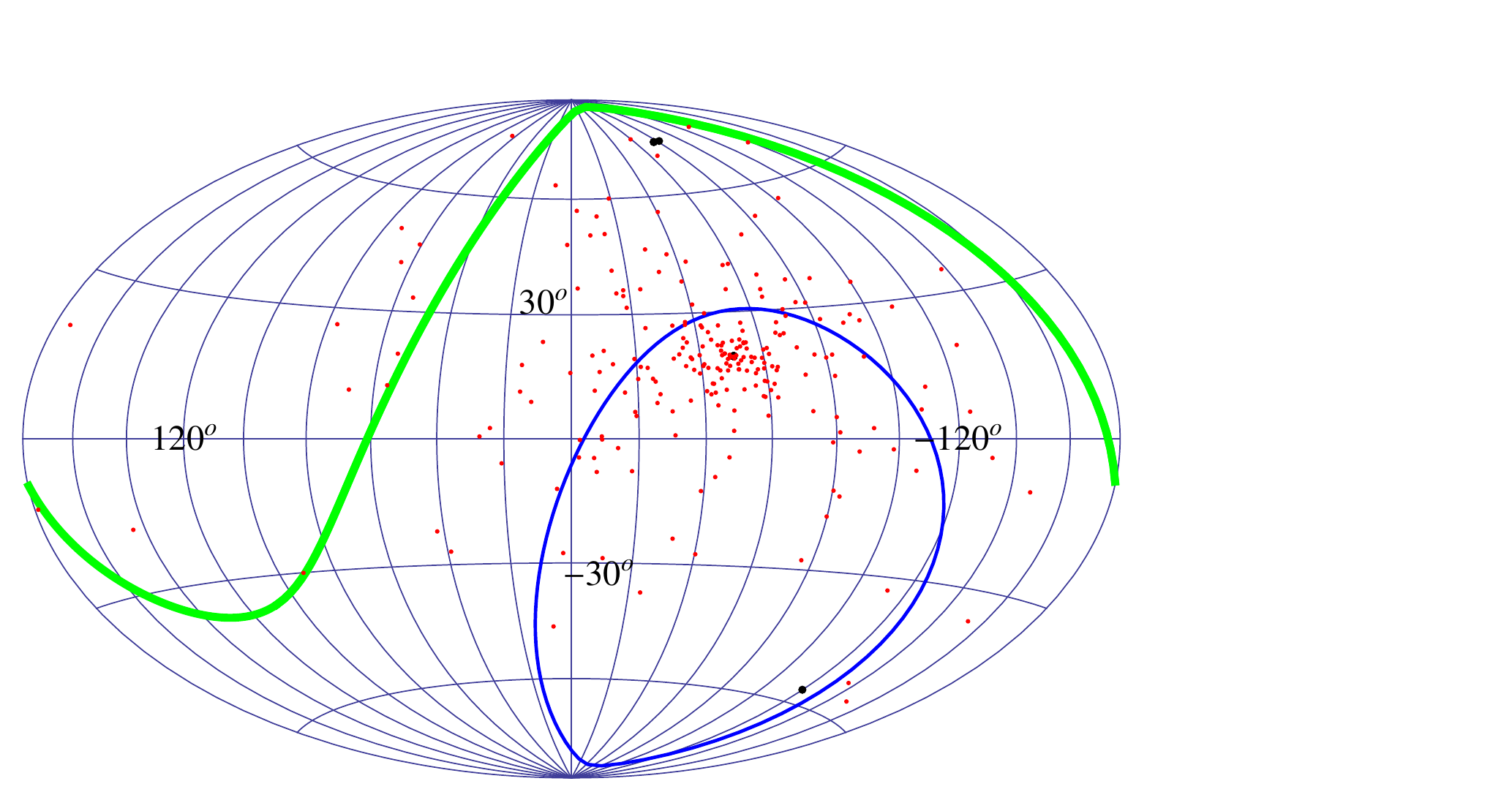}
\caption{Aitoff projection in galactic coordinate of the scattered 300 events (red dots) coming from a selected population of AGN sources (black dots). The model for particle acceleration is spin-down and the main contributors are NGC 5128, M87 and NGC 1316 in this order. Above the thick line is the area from the sky not seen by Auger (declinations above than 24.8 degrees) and the thin line surrounds the area from the sky not seen by the HiRes experiment (declinations less then -32 degrees).}
\label{AGNScatteringspindown}
\end{figure}

\begin{figure}[htpb]
\centering
\includegraphics[viewport=0cm 0cm 21cm 11cm,clip,scale=0.8]{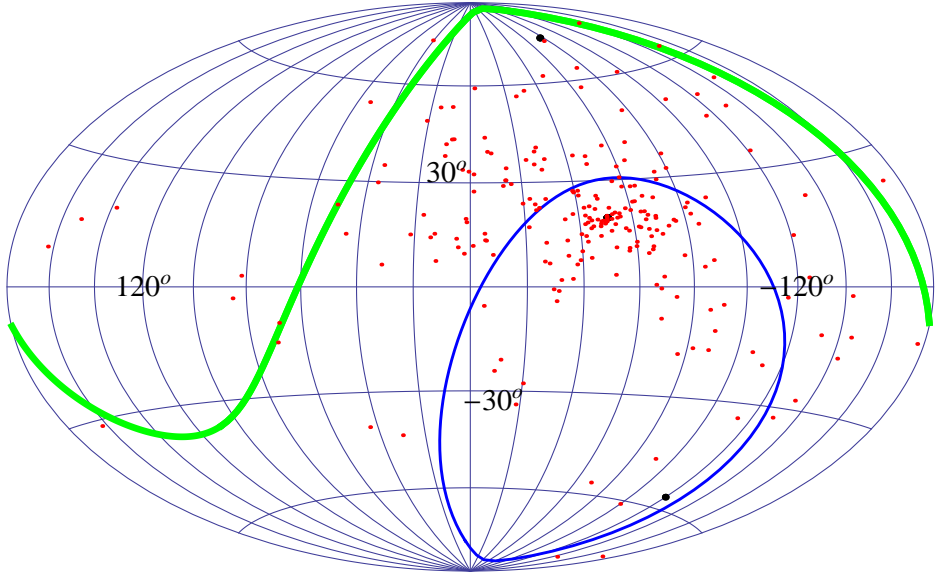}
\caption{Aitoff projection in galactic coordinate of the scattered 300 events (red dots) coming from a selected population of AGN sources (black dots). The model for particle acceleration is accretion dominated and main contributors are NGC 5128, M87 and NGC 1316 in this order. Above the thick line is the area from the sky not seen by Auger (declinations above than 24.8 degrees) and the thin line surrounds the area from the sky not seen by the HiRes experiment (declinations less then -32 degrees).}
\label{AGNScatteringaccretion}
\end{figure}

Regarding the two models for calculation the cosmic rays fluxes, spin-down and accretion dominated, Fig.~\ref{AGNScatteringspindown} and Fig.~\ref{AGNScatteringaccretion} shows that we get a strong clustering around the main source that gives most of the events in both cases, this can give a clear indication of the real sources in the sky.

\clearpage

We performed three separate runs of analysis regarding the simulated events, (Table~\ref{tableHIAUAsimetry}) counting how many events an experiment of cosmic rays in the Northern hemisphere of the Earth, like HiRes will be able to detect and how many events an experiment of cosmic rays from the Southern part of the Earth, like Pierre Auger Observatory can record. We get a strong asymmetry between North and South sometimes more then 2 times more events in the Pierre Auger Observatory sky than HiRes one for the AGN two types of acceleration mechanism and a weak asymmetry between North and South in the case of GRB.

Here we observe that the GRB model systematically favors the Northern sky, while the radio galaxy model favors the Southern sky.  The differences are not very big in the GRB case, but highly significant in the radio galaxy case.

\begin{table}[h!]
\begin{center}
\small
\caption{HiRes vs Auger asymmetry between arrival direction from two type of sources
and in the case of AGN two types of acceleration mechanism.}
\begin{tabular}{|c|c|c|c|}
\hline
Source&Count HvsA Run1&Count HvsA Run2&Count HvsA Run3\\
\hline
GRB&	151/108&	148/127&	148/119\\
\hline
AGN spindown&	91/224&	107/220&	92/222\\
\hline
AGN accretion&	93/215&	86/226&	85/222\\
\hline
\end{tabular}
\label{tableHIAUAsimetry}
\end{center}
\end{table}

\begin{figure}[htpb]
\centering
\includegraphics[viewport=0cm 0cm 21cm 11cm,clip,scale=0.8]{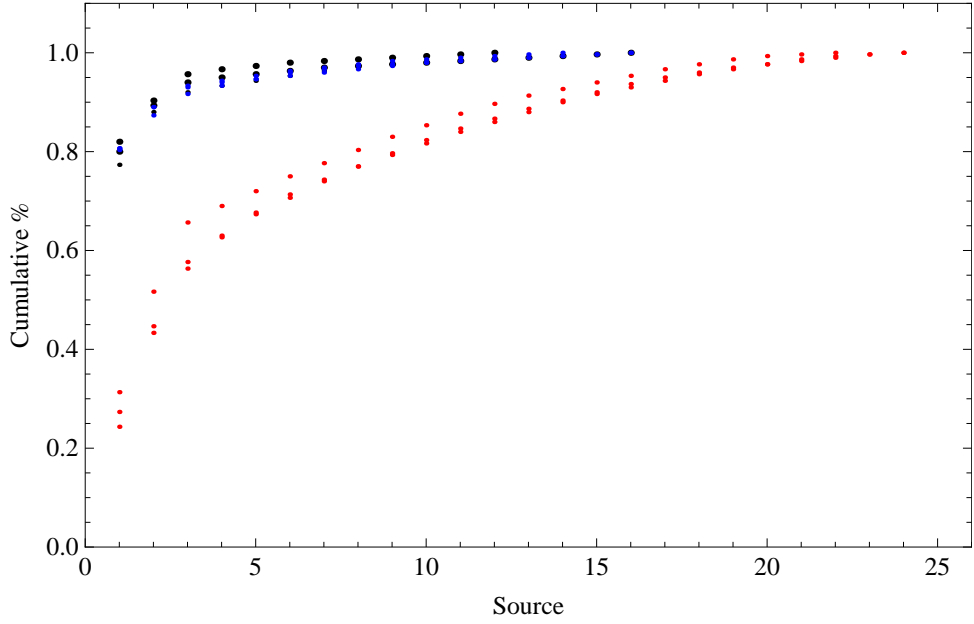}
\caption{Contribution of sources for ultra-high energy cosmic rays. Red is for GRB sources (the lower set of points), black for AGN sources in spin-down model and blue is for AGN sources in accretion mode (nearly coincident upper sets of points, with the black ones slightly higher than the blue ones). Each time three runs were performed.}
\label{ContributionTable}
\end{figure}

In Fig.~\ref{ContributionTable} we show that in the case of GRB many sources contribute (10 sources give 80\%) to give the distribution of events and in the case of AGN just a few of them (one source gives 80\%) give the majority of events, dominating the whole sky in cosmic rays.

The full statistical analysis covering $10^{6}$ Monte-Carlo runs for different galactic and extragalactic magnetic field models including different particle acceleration models, also taking into account the sky sensitivity for the two cosmic rays experiments, Pierre Auger Observatory and HiRes, will be available in a following paper \citep{2011Curutiu}. We only note that our conclusions drawn from only 3 runs are confirmed by all those statistically large simulations.

All this just reflects the well-known fact that the sky is not homogeneous in the nearby Universe.

\subsection{Statistical approach without redshift limit}

It is important to check with the known radio luminosity function, whether low power RGs dominate in the predicted cosmic ray power, or the high power RGs are the dominant ones.  If it were possible, that low power RGs dominate in the sum, then any scattering argument might be superfluous.  On the other hand, if only high power RGs contribute, then we can concentrate on those few, and the scattering argument will be important.

Here we use the radio luminosity function as shown in, e.g. \citet{2001MNRAS.322..536W}, this luminosity function has a slope of -1.6.  From \citet{1991Natur.349..138R,1997ApJ...477..560E} we have used already above, that the jet-power relates to the total radio luminosity as a power-law.

\begin{equation}
L_{jet} \, \sim \, L_{rad}^{\alpha}
\end{equation}

with ${\alpha}$ in the range 0.7 to 1.0.

This implies, recalling from above the various dependencies for the spin-down mode and for the accretion mode, that the ultra high energy cosmic ray luminosity $L_{UHECR}$ is

\begin{equation}
L_{UHECR, spd} \, \sim \, L_{jet} \, \sim \, L_{rad}^{\alpha}
\end{equation}

in the spin-down mode, and

\begin{equation}
L_{UHECR, acc} \, \sim \, L_{rad}^{2/3}
\end{equation}

in the accretion mode. This exponent is very close to the lower end of the allowed range for the spin-down mode.

The maximal particle energy then is

\begin{equation}
E_{UHECR, spd} \, \sim \, L_{jet}^{1/2} \, \sim \, L_{rad}^{\alpha/2}
\end{equation}

in the spin-down mode, and

\begin{equation}
E_{UHECR, acc} \, \sim \, L_{jet}^{1/2} \, \sim \, L_{rad}^{\alpha/2}
\end{equation}

in the accretion mode. However, since in the accretion mode the accretion and jet power is not very far from a fixed proportion of order 0.1 to the Eddington limit, we can rewrite this latter relationship as

\begin{equation}
E_{UHECR, acc} \, \sim \, L_{rad}^{3 \alpha/2}
\end{equation}

Next we have to work out the real numbers to obtain the realistic contribution, and also to test, whether any radio galaxy can provide enough energy for the particles.

In either mode the maximal particle energy is limited by the Poynting flux limit, which gives

\begin{equation}
L_{jet} \; = \, 10^{47} \, {\rm erg/s} \, f_{flare} \, {\left(\frac{E_{max}}{Z \, \gamma_{sh} \, 10^{21} \, {\rm eV}}\right)}^{2}
\end{equation}

where $f_{flare}$ is a factor allowing for flaring, so that what we observe today in radio luminosity may not relate properly to the time integral of the emission of ultra high energy particles, implying that $f_{flare}$ could be very different from unity.  We also include here a factor from the possibility of a relativistic shock $\gamma_{sh} \, \geq \,  1$ \citep{1999MNRAS.305L...6G}.  The jet in Cen A has a power
 of the order of $10^{43}$ erg/s, and the jet in M87 a power of the order
 of $10^{45}$ erg/s, both of them failing the test to allow for particles
 with energy as high as observed (\citet{2003NewAR..47..219W}).
 If we consider flaring to give a factor of 10, and also consider that no
 particles are confirmed to be above $3 \times 10^{20}$ eV, then M87 is
 allowed.  However, Cen A is definitively excluded, without allowing
 for $Z \, \gamma_{sh} \simeq 10$, strongly supporting a the point view
 that the observed particles might be heavy nuclei.

The minimum power defines a starting radio luminosity for integrating the various contributions:

M87 is basically the minimum, if we allow only for $Z \, \gamma_{sh} \simeq 1$ and flaring, and Cen A is the minimum, if we allow for $Z \, \gamma_{sh} \simeq 10$, and also flaring.  This limit is independent of the black hole mass, since the Poynting flux is a clear low limit to the energy flow along the jet \citep{1976Natur.262..649L,1995A&A...293..665F}.
 In the graph of the radio luminosity function in \citet{2001MNRAS.322..536W},
 both M87 and Cen A are outside the graph by a factor of 5 to 100 in
 luminosity, and based on other publications by the same authors we assume,
 that the radio luminosity function continues with the same slope.
 We note, however, that the extended radio luminosity of Cen A
 has been shown to be considerably larger than previously observed
 (\citet{1993A&A...269...29J}), and so we have to assume that this is
 also true at lower radio frequency, diminishing the factor of
 discrepancy for Cen A to about 20.

In order to determine, which radio power dominates the integral, we then take the slope of the radio luminosity function, -1.6, and integrate:

\begin{equation}
\int L_{rad}^{-1.6} \, L_{UHECR} \, d L_{rad} \, \sim \, \int L_{rad}^{-1.6} \, L_{rad}^{\alpha} \, d L_{rad}
\end{equation}

where $\alpha$ is 2/3, or 0.7 to 1.0.  The factor $\alpha - 0.6 \, > 0$ determines by how much the high power RGs dominate, since $\alpha$ can range from 2/3 to 1.0, this is not by much.  Considering the graph in \citet{1997ApJ...477..560E}, it appears that for low power RGs, $\alpha \, \simeq \, 1.0$ is a better approximation, connecting them to other RGs. This suggest that the real exponent is 0.4 in the integral above and so, high power RGs dominate among the FR-I RGs.  This is in contrast to the arguments in \citet{1995A&A...293..665F}, which were developed for accretion dominated jets and powerful RGs, and suggested a value for $\alpha$ of 2/3.

It might be worth considering a caveat here:  If we could convince ourselves, that the exponent for low power RGs could be 2/3 in the connection between total radio luminosity and jet power, then all RGs would equally contribute to the flux of cosmic rays.  However, the maximal energy is another condition, and that does not allow most of these RGs to contribute above $10^{20}$ eV, so setting a sharp lower bound.

Therefore we conclude tentatively, that for low power RGs

\begin{equation}
F_{UHECR, int} \, \sim \, L_{rad, max}^{0.4},
\end{equation}

where $L_{rad, max}$ is determined from the maximum radio luminosity allowed within the volume accessible for the propagation of ultra high energy cosmic rays.  Considering a sphere with a radius of 100 Mpc as the volume accessible, both from particle interaction with the microwave background \citet{2008JCAP...10..033A} and magnetic scattering \citet{2008ApJ...682...29D}, then this implies a source density of $10^{-6.4} \, Mpc^{-3}$, which is reached for luminosity of order $10^{32.3}$ erg/s/Hz at 151 MHz, corresponding to $10^{31.1}$ erg/s/Hz at 5 GHz (using a spectral index of -0.8).  However, allowing for flaring by a factor of 10 implies that only 0.1 of all sources contribute significantly, and so we require a density with a factor of 10 higher, implying a maximum radio power of $10^{30}$ erg/s/Hz at 151 MHz, corresponding to $10^{28.8}$ erg/s/Hz at 5 GHz, a factor of 10 outside the plot in \citet{2001MNRAS.322..536W}.  Therefore M87 may be the only source in the reachable volume to allow protons, and Cen A may be one of the very few sources to allow only heavies. If we consider the FIR/radio ratio as an indicator for an accompanying starburst, then the radio galaxies N4651, I5063 and N5793 are all above Cen A in this measure, but they all are below 3 percent of Cen A in their possible particle flux contribution, even in the sum below 5 percent.  So Cen A seems unique, and our location in the nearby universe seems special, defying the Copernican principle that nothing should be special about our location in the universe.

Therefore we conclude that Cen A and M87 are the best radio galaxy candidates to produce ultra high energy cosmic ray particles, Cen A only heavy nuclei, and M87 mostly light nuclei.

As a check with the available data we constructed the integral distribution of cosmic rays fluxes for the two cases of particle acceleration models.

\begin{figure}[htpb]
\centering
\includegraphics[viewport=0cm 0cm 20cm 10.5cm,clip,scale=0.7]{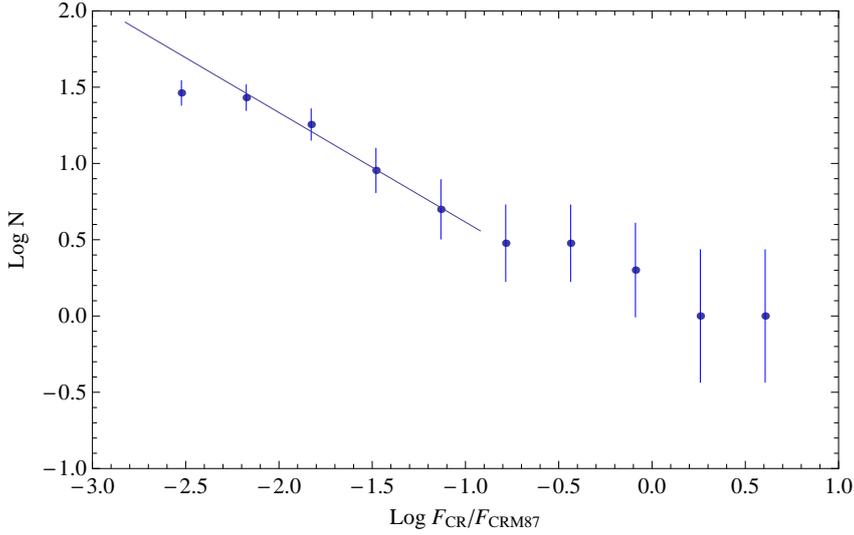}
\caption{Integral distribution of cosmic rays flux in the case of spin-down acceleration model.}
\label{SpinDownDistrib}
\end{figure}

In the case of the spin-down model we ask the question if the small sources can contribute to the cosmic ray flux in a significant way, at least at the level of the most important sources. Taking into account that as we approach the limit of small cosmic ray flux we are no longer complete, we exclude the first point of the integral distribution plot. We consider the interval between $10^{-3}$ to $10^{-1}$ of $F_{CR}/F_{CRM87}$ and make a power low fit of the points in this interval:

$Log N = -0.1(\pm 0.08) - 0.71(\pm 0.05) Log F_{CR}/F_{CRM87}$.

In this interval, we get around 122 sources that will contribute.
Using the differential plot for the same interval we obtain again a fit like:

$Log N = 1.78(\pm 0.12) + 0.43(\pm 0.05) Log F_{CR}/F_{CRM87}$.

If we sum all of the contribution from each of this source, using the fit from the differential plot and taking into account the logarithmic counting that we used, we get a total up to a level to $0.69$ of $F_{CR}/F_{CRM87}$. If we compare only with the contribution from Cen A and M87 itself, which is around 10 $F_{CR}/F_{CRM87}$, this is very low, so the important contributors dominate the sky.

\begin{figure}[htpb]
\centering
\includegraphics[viewport=0cm 0cm 20cm 10.5cm,clip,scale=0.7]{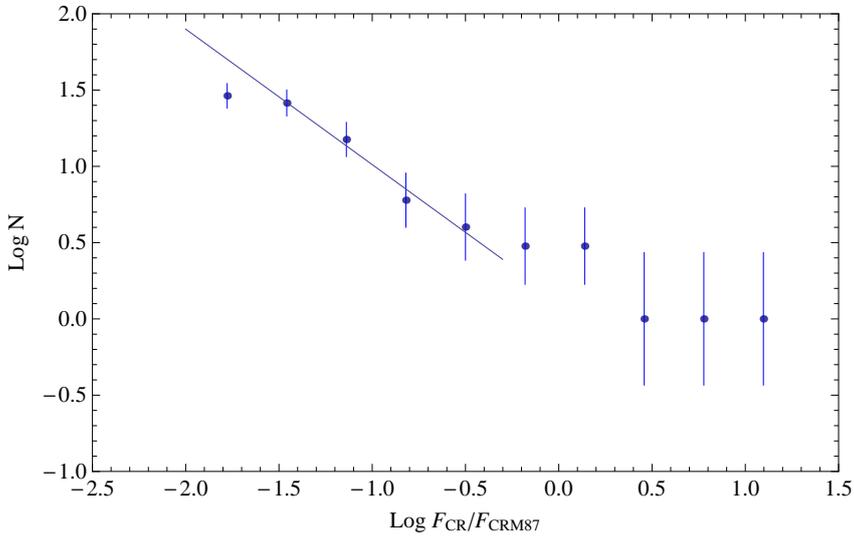}
\caption{Integral distribution of cosmic rays flux in the case of accretion dominated acceleration model.}
\label{SpinDownDistrib}
\end{figure}

Using the same procedure as for the case of spin-down model, we fit the integral distribution of cosmic rays for the accretion dominated acceleration model. Considering the same interval of flux ratio, between $10^{-3}$ to $10^{-1}$ of $F_{CR}/F_{CRM87}$ we get:

$Log N = 0.12(\pm 0.09) - 0.88(\pm 0.08) Log F_{CR}/F_{CRM87}$.

The number of sources that can contribute to give the flux can be readily obtained from this and is around 613 sources for the interval under study.
Using the differential plot of the distribution we find a fit like:

$Log N = 2.08(\pm 0.63) + 0.62(\pm 0.15) Log F_{CR}/F_{CRM87}$.

Again, if we sum the contribution from each source as before, we get a total of $13.2$ of $F_{CR}/F_{CRM87}$. Comparing with the flux ratio from Cen A and M87, which is around 13 $F_{CR}/F_{CRM87}$, we conclude that they are quite similar and so, the small sources are at the same level as the principal contributors of cosmic rays if we consider the most favorable distribution for the small sources. However, the maximal particle energy condition eliminates all these low power sources:  In the sum as a function of energy this could lead to a much steeper spectrum than any individual spectrum, as suggested some time ago by \citet{2006PhRvD..74d3005B}.

For these two models of acceleration and for the case when the number of small sources follows a power law, the flux of cosmic rays is given in principle by few very powerful, close sources, as Cen A and M87 and is not dominated by the small sources.
We derived this result under maximal assumptions, not taking into the magnetic scattering nor any interactions, so as to be independent of any assumptions on chemical compositions, which would strongly influence both. We started by using a complete sample of identified radio galaxies with a redshift limit, then progressed using a radio luminosity function without any redshift limit. Finally we note that in \citet{2011arXiv1106.0625B}, it was shown that the observations do not actually require even a single second extragalactic source to explain the UHECR spectrum between $10^{15}$ eV and $10^{21}$ eV to within the observational errors; the entire spectrum can be matched with a model taken from \citet{2003PhRvD..68j3004S} and using an approach about the effect of a relativistic shock taken from \citet{2001MNRAS.328..393A}, shifting a spectrum of local galactic cosmic rays to higher particle energies. The spectrum can be matched by using the radio galaxy Cen A and our own Galaxy, consistent with \citet{1963SvA.....7..357G} and what we find here, that Cen A is expected to dominate by 10:1 over the second source, and all the weaker sources contribute negligibly, even when integrating over the weak fluxes.

\subsection{High energy neutrinos}

Obviously, as soon as one accelerates ultra high energy protons and nuclei, these particles undergo interactions and produce neutrinos, see \citet{2009APh....31..138B}.  As these interactions happen much closer to the central engine than the acceleration of the ultra high energy particles, which we observe, the best selection for such sources is a sample of flat spectrum radio sources.  Here is a  complete sample for this discussion in Table~\ref{tableneutrino}.
We show our third complete sample, here of flat spectrum radio sources, flux density limited at 5 GHz again, but with no redshift limit.  These are normally interpreted to be relatively low power RGs, whose relativistic jet points towards Earth \citep{1988A&A...206..245W}.  We give the name, the redshift, the 2.7 GHz flux density, the type of host galaxy and AGN, and the Galactic coordinates.  These AGN are expected to produce ultra high energy cosmic ray particles, which are only recognizable to us through the process of interaction, when they produce neutrinos or high energy photons, e.g. \citet{2010arXiv1012.0204B}.

The source with the highest flux density may also be the source with the highest neutrino flux, and so we ordered this table by flux density at 2,7 GHz. However, AMANDA and IceCube are the most sensitive near or above the celestial equator, its horizon at the South pole, and we mark those sources, like 3C279, with an asterisk. Therefore we predict these six sources to be prime candidates as neutrino sources at high energy, from interactions of ultra high energy particles with the surrounding photon field, near the first strong shock in their relativistic jet. It is not a priori clear, whether the beaming in radio emission is narrower or wider than the predicted beam in neutrinos. If it is narrower, then the subset of the sources with strong variability would again be preferred, like 3C273, 3C279, and 0440-003. Therefore we predict these sources to be good candidates to be discovered in high energy neutrinos. If we go by radio flux density, then clearly 3C273 and 3C279 are probably the best candidates as high radio flux density and should correlate very well with the futured observed high fluxes of UHE-neutrino.

\begin{table}
\begin{center}
\small
\caption{Properties of the top 25 flat and inverted radio spectrum sources, highest in flux density at 2.7 GHz.}
\begin{tabular}{|c|c|c|c|}
\hline\hline
Name&Redshift&Flux density&Type\\
&&Jy&\\
\hline
3C 273$*$&0.158&38.9&blazar; Sy1    LPQ\\
ESO 362- G 021&0.061042&12.5&N galaxy;HPQ   BLLAC\\
3C 279$*$&0.5362&11.2&blazar;HPQ     BLLAC\\
3C 454.3&0.859&10&blazar         HPQ\\
NGC 1275&0.0175&9.64&cD;pec;NLRG\\
$[$HB89$]$ 2134+004$*$ &1.932&7.6&Opt.var.       LPQ\\
2MFGC 06756&0.24117&7.54&Radio galaxy   Sy3\\
$[$HB89$]$ 1127-145&1.184&6.5&blazar         LPQ\\
$[$HB89$]$ 0438-436&2.863&6.2&HPQ\\
3C 345&0.5928&6.08&Opt.var.       HPQ\\
BL Lac &0.0686&5.21&Opt.var.       BLLAC\\
$[$HB89$]$ 2203-188 NED02&0.6185&5.2&blazar         LPQ\\
$[$HB89$]$ 0923+392&0.69528&4.6&Opt.var.; Sy1   LPQ\\
$[$HB89$]$ 0637-752&0.653&4.51&FSRQ           Sy1\\
3C 446$*$&1.404&4.4&Opt.var.;HPQ   BLLAC\\
$[$HB89$]$ 0834-201&2.752&4.15&blazar         LPQ\\
$[$HB89$]$ 2345-167&0.6&4.08&Opt.var.;BLLAC HPQ\\
PKS 1549-79&0.1501&4.02&Sy2\\
ABELL S0463&0.0394&3.84&I-II   $[$BM$]$;R: $[$A$]$\\
$[$HB89$]$ 0537-441 &0.894&3.84&blazar;HPQ     BLLAC\\
4C +12.50&0.12174&3.8&S0;Double nuc. Sy2\\
PKS 0742+10$*$&2.624&3.74&FSRQ\\
$[$HB89$]$ 0440-003$*$ &0.844&3.73&Opt.var.;HPQ  blazar\\
$[$HB89$]$ 0208-512&0.999&3.56&blazar;HPQ     BLLAC\\
OJ +287&0.306&3.42&Opt.var.       BLLAC\\

\hline
\end{tabular}
\label{tableneutrino}
\end{center}
\end{table}

\section{Conclusion}

We approach the question of the origin of ultra high energy cosmic rays by using complete well-defined samples of sources. We consider first starburst galaxies where the Northern and Southern sky have a large probability to have events, with the Northern sky slightly more populated. However, since GRB from galaxies are rare, and of course in each individual case there are extended periods with no GRB, the statistics allows different possible scenarios concerning the contribution from each galaxy.  Furthermore, GRB can also allow heavy nuclei, in which case many conclusions, apart from the North-South difference are similar to the radio galaxy hypothesis. The GRB hypothesis easily allows for many galaxies to contribute rather than few as is the case for RGs.  Finally, mixed scenarios are also conceivable, in which we have a starburst phase accompanying a galaxy merger, this will lead to a merge of the central super-massive black holes associated with each galaxies giving at the end a radio galaxy with a starburst.  In this case the direct acceleration of particles from GRB could compete with the acceleration involved in radio galaxy caused by particle  seeds from Wolf-Rayet star explosions. In the case of Cen A we note that the central relativistic jet is predicted to produce more UHECR than the relatively weak starburst through its GRBs.

Therefore there are four simple conclusions for the hypothesis
 that ultra high energy particles come from GRB:

 a)  The GRB rate per star formation rate must be at least an order of
 magnitude larger than assumed previously
 \citep{2000A&A...358..409P,2001AAS...199.1202P}.  The selection effects
 are so large, that this may be possible.

 b)  As starburst galaxies are mostly in the North, more particles should
 arrive in the North, to the degree that magnetic scattering does not
 obscure all original source directions.  The nearby starburst in Cen A
 would also produce a cloud of events coming from that direction, but the
 Cen A starburst is not especially strong compared to well known starburst
 galaxies at a similar distance such as M82. This implies that in the case
 of GRB, to have a dominating single source such as Cen A in UHECR
 would be hard to understand.  However, the Southern sky would still
 contain the second largest contribution, since NGC4945 is not far from
 Cen A. NGC253 is near the Galactic South pole, and could show up as a
 separate cloud of events.  Therefore, in the case of GRB we have several
 likely sources of relatively similar possible contributions.

 c)  We also have the conclusion, that the turn-off in the spectrum may
 be due to either the source or the microwave background. In the first case,
 the shape of the turnoff is given by the rapid source evolution after
 becoming optically thin, and in the second case the shape should have
 the standard GZK-spectral turnoff, convolved with the source spectrum.

 d)  Almost all particles could be protons except for the very high energy particles that may still be neutrons which will point straight back to the source. But we have to add that this is just a possibility offered by GRB, not a foregone conclusion. This option can easily be tested with starburst galaxies such as M82 and events at very high energy, since a few percent of them would survive as neutrons, and so would point straight back at the origin.

Using a complete sample of steep spectrum RGs, so neither relativistic jets pointed at the observer nor starburst galaxies, we estimate the total flux of ultra high energy particles, that might reach Earth.  We also estimate their maximal particle energy. There are few sources capable of providing the particle energies and the flux required by observations.  On the positive side we find, that Cen A is expected to be the dominant source, however, only if it accelerates heavy nuclei.  This is doable, if the injection particles are the heavy nuclei near PeV energies in the knee region of normal cosmic rays, accelerated in the starburst currently occurring in Cen A.  M87 and a few other RGs are also viable sources, but we see no argument why they should produce more than light nuclei. However, we have to note that no experiment observes any clustering of events around M87 on the sky.

There are two predictions as to how to distinguish these two sites of origin:

In the GRB picture all events could be just protons, we need even larger magnetic fields than in \citet{2008ApJ...682...29D} to scatter them even at high energy to near isotropy. And, there might be more events in the North than in the South. We have only a qualitative proposal as to how to model the observed spectrum.

Since it has been shown that GRB are capable of giving heavy nuclei in UHECR at the same high energies as RGs, then the only distinction to RGs with starburst galaxies would be the very different scaling. In the GRB case the flux would relate to the strength of the starburst, so the mass of gas turned into stars per unit of time, and in the radio galaxy case the flux would relate to the power of the jet. Also, the explanation of the spectral downturn would have to be very different. Another aspect is that the number of contributing sources would be large for GRB and their host starburst galaxies and very small for RGs. High energy neutrino observations have the power to decide on whether RGs are sources of ultra high energy cosmic rays, since in that case, flat spectrum radio sources, RGs pointed towards us, should be detected in neutrinos. In the opposite case, high energy neutrino pointing straight back at starburst galaxies would be a signature.

In the radio galaxy picture, Cen A is by far the most probable source, but that source requires heavy nuclei. The observed spectrum can be readily modeled, \citet{2010ApJ...720L.155G}, using the spectrum of cosmic rays across the energy range near and beyond $10^{15}$ eV as seeds, \citet{1993A&A...274..902S}, and then one time shift suggested by \citet{1999MNRAS.305L...6G} to high energy given by a relativistic shock. In such a case, the magnetic field between Cen A and us can be modeled as in \citet{2008ApJ...682...29D}. Another consequence of the RGs as sources is that there are more events in the South than in the North, as the distribution of sources is isotropic on the sky.

The determination of the chemical composition will allow a final judgment on which of these two pictures is a viable candidate to explain the data, or if both of them fail. As noted in \citet{2010ApJ...720L.155G}, there could be telling differences between North and South in the apparent chemical composition, if so, it would allow the sites of origin to be well circumscribed. Furthermore, in the radio galaxy picture flat spectrum radio sources visible form IceCube such as 3C273, 3C279, or 3V454.3 would be strong sources of high energy neutrinos \citep{1992A&A...260L...1M}.

\section{Acknowledgements}

Work with PLB was supported by contract AUGER 05 CU 5PD 1/2 via DESY/BMB and by VIHKOS via FZ Karlsruhe; by Erasmus/Sokrates EU-contracts with the universities in Bucharest, Cluj-Napoca, Budapest, Szeged, Cracow, and Ljubljana; by the DFG, the DAAD and the Humboldt Foundation; also by research foundations in Korea, China, Australia, India and Brazil. Support for TS comes from DOE grant UD-FG02-91ER40626. L.I.C. was supported by PN-II-RU-TE-2011-3-018 and CNCSIS Contract 539/2009. The authors would like to acknowledge discussions of these questions with many colleagues, but especially with J. Becker, J. Bluemer, A. Caramete, R. Engel, T. Gaisser, L. Gergely, Gopal-Krishna, A. Haungs, K.H. Kampert, L. Popa, V. de Souza, Ch. Spiering, and P. Wiita. PLB acknowledges discussions with his Pierre Auger Observatory collaborators, especially H. Glass. This research has made use of the NASA/IPAC Extragalactic Database (NED) which is operated by the Jet Propulsion Laboratory, California Institute of Technology, under con- tract with the National Aeronautics and Space Administration. This research also made use of the ViZier system at the Centre de Donnee ́s astronomiques de Strasbourg (CDS) \citep{2000A&AS..143...23O}.

\bibliographystyle{aa} 
\bibliography{references} 

\end{document}